\documentclass{article}

\usepackage{hyperref}
\usepackage{amssymb} 
\usepackage{amsmath}
\usepackage{amsthm}
\usepackage{dsfont}
\usepackage{tikz}
\usepackage{enumitem}   
\usetikzlibrary{arrows}
\usepackage{algorithm}
\usepackage{algpseudocode}
\usepackage{hyperref}
\usepackage[numbers]{natbib}
\usepackage{comment}
\usepackage{color}
\specialcomment{Notes}{%
  \begingroup\color{blue}}{%
  \endgroup}
\allowdisplaybreaks

\RequirePackage{amsthm}
\RequirePackage{thmtools}
\declaretheorem[title=Theorem,numberwithin = section, style=plain]{thm}
\declaretheorem[numberlike=thm, title=Lemma, style=plain]{lem}
\declaretheorem[numberlike=thm, title=Corollary, style=plain]{cor}
\declaretheorem[numberlike=thm, title=Definition, style=plain]{defn}
\declaretheorem[numberlike=thm, title=Proposition, style=plain]{prop}
\declaretheorem[numberlike=thm, title=Assumption, style=plain]{assume}
\declaretheorem[title=Remark, style=remark]{rem}
\declaretheorem[title=Example, style=remark]{exa}

\DeclareMathOperator{\argmin}{argmin}
\DeclareMathOperator{\diag}{diag}
\newcommand{\N}{\mathbb{N}}
\newcommand{\R}{\mathbb{R}}
\newcommand{\myE}[2]{\mathbb{E}_{#1}\left[ #2 \right]}

\begin{document}

\author{Lukas Gonon$^{1,2}$ \and  Thilo Meyer-Brandis$^3$ %\thanks{
%Department of Mathematics, LMU Munich, Germany. {\tt meyerbra@math.lmu}, {\tt weber@math.lmu.de}} 
\and Niklas Weber$^3$ 
%\samethanks[2]
}
 
\title{\textbf{Computing Systemic Risk Measures with Graph Neural Networks}}
\date{September 9, 2025}
\maketitle

\begin{abstract}
This paper investigates systemic risk measures for stochastic financial networks of explicitly modelled bilateral liabilities. 
We extend the notion of systemic risk measures based on random allocations from Biagini, Fouque, Fritelli and Meyer-Brandis (2019) to graph structured data. 
In particular, we focus on aggregation functions that are derived from a market clearing algorithm proposed by Eisenberg and Noe (2001). In this setting, we show the existence of an optimal random allocation that distributes the overall minimal bailout capital and secures the network.\\
We then study numerical methods for the approximation of systemic risk and optimal random allocations.
Our proposition is to use permutation-equivariant architectures of neural networks such as graph neural networks (GNNs) and a class that we name (extended) permutation-equivariant neural networks ((X)PENNs).
The performance of these architectures is benchmarked against several alternative allocation methods. 
The main feature of GNNs and (X)PENNs is that they are permutation-equivariant with respect to the underlying graph data. In numerical experiments, we find evidence that these permutation-equivariant methods are superior to other approaches.
\end{abstract}

\bigskip

\textbf{Keywords:} systemic risk measure, financial network, contagion, default, graph neural network, permutation-equivariant, universal approximation, message passing, deep learning
 \\
\textbf{Mathematics Subject Classification (2020):} 68T07, 91G45, 91G60, 91G70\\
\textbf{JEL Classification:} C61, G10, G21, G32, G33\\

\makeatletter
\addtocounter{footnote}{1} \footnotetext{%
Department of Mathematics, Imperial College London, United Kingdom. {\tt l.gonon@imperial.ac.uk}}
\addtocounter{footnote}{1} \footnotetext{
School of Computer Science, University of St. Gallen, Switzerland.  {\tt{lukas.gonon@unisg.ch} }}
\addtocounter{footnote}{1} \footnotetext{
Department of Mathematics, LMU Munich, Germany. {\tt meyerbra@math.lmu}, {\tt weber@math.lmu.de}}
\makeatother

% ###################################################################

%###################################################
\section{Introduction}

Risk measures are fundamental tools in financial mathematics that provide a way to quantify and manage the risk associated with financial positions or portfolios. The development of these measures has evolved significantly over time, driven by both theoretical advancements and practical needs in the financial industry. The early developments in risk measures can be traced back to Harry Markowitz's seminal work, ``Portfolio Selection'', in 1952 \cite{markowitz1952portfolio}, introducing the mean-variance framework. In the 1990s, the Value at Risk (VaR), most notably developed by JP Morgan, became widely used in risk management and regulatory frameworks due to its simplicity and intuitive appeal.
However, its limitations led to a line of literature developing theoretically sound concepts of risk measures, such as coherent, convex, or monetary risk measures; see for example \cite{artz}, \cite{follmer2002convex}, \cite{frittelli2002putting}, \cite{foell}.

As the financial system became more interconnected and complex, in particular in the aftermath of the financial crisis 2007–2008, the focus widened towards understanding and measuring systemic risk, i.e., the risk that a shock to the financial system can cause a collapse of significant parts of the system; see, for example, \cite{acharya2017measuring}, \cite{adrian2016covar}.

Many of these systemic risk measures can be described as the application of a univariate risk measure $\eta: L^0 \to \R$ to some aggregated systemic risk factor $\Lambda(X)$, where $X = (X_1,\ldots,X_N) \in L^0(\R^N)$ represents random risk factors and $\Lambda:\R^N \to \R$ is a function that aggregates the individual risk factors $X_1,\ldots,X_N$ to a one dimensional systemic risk factor, such that the systemic risk is given by
\begin{align*}
    \rho(X) = \eta(\Lambda(X)).
\end{align*}
If $\eta$ is cash-invariant (see \cite{foell}), the systemic risk can often be interpreted as the amount of cash that is needed to secure the position in terms of risk measure $\eta$. If $\mathbb{A}$ is the corresponding acceptance set of $\eta$, the systemic risk measure can be reformulated as

\begin{align}\label{basic_sys_risk_after_agg}
    \rho(X) = \inf\{m \in \R \mid \Lambda(X) + m \in \mathbb{A}\}.
\end{align}

In \cite{chen2013axiomatic}, an axiomatic approach for such systemic risk measures and conditions on when they can be decomposed into aggregation function and univariate risk measure is provided. This setting is further investigated for example in \cite{hoffmann2016risk}, \cite{kromer2016systemic}.

The formulation in (\ref{basic_sys_risk_after_agg}) highlights that the amount of cash $m \in \R$ is added to the \textit{aggregated} position $\Lambda(X)$. In contrast to this setting, where the system is secured \textit{after aggregation}, an alternative approach investigates the setting where capital is allocated \textit{before aggregation}. This is motivated, for example, by aggregation functions that capture financial contagion between the institutions. In this case, it can be significantly cheaper to prevent contagion by allocating capital \textit{before aggregation} as opposed to securing the system \textit{after aggregation} when the contagion has already occurred. This idea yields systemic risk measures of the form
\begin{align}\label{line:sys_risk_vec_det_alloc}
    \rho(X) = \inf \left\{\sum_{i=1}^N m_i \,\middle|\, m = (m_1,\ldots,m_N) \in \R^N, \Lambda(X + m) \in \mathbb{A} \right\}.
\end{align}

Systemic risk measures based on allocating capital \textit{before aggregation} have been investigated in the context of set-valued risk measures and \textit{deterministic} allocations in \cite{feinstein2017measures}. In this approach, each acceptable vector $m=(m_1,\ldots,m_N)$ can be interpreted as a possible choice of capital requirements for the institutions $(X_1,\ldots,X_N)$ such that the resulting financial system is safe after aggregation, i.e.,\ $\Lambda(X+m) \in \mathbb{A}.$ Another approach, which we focus on in this article, was developed in \cite{biagini2019axiom} and \cite{biagini2020fairness}. Here, the \textit{deterministic} capital allocation $m=(m_1,\ldots,m_N) \in \R^N$ is extended to a \textit{random} capital allocation $Y \in L^0(\R^N)$. Then, the systemic risk measure is defined as
\begin{align}\label{line:sys_risk_vec_rand_alloc}
    \rho(X) = \inf_{ Y \in \mathcal{C}} \left \{\sum_{i=1}^N Y_i \,\middle|\, \Lambda(X+Y)\in \mathbb{A}\right\},
\end{align}
where the set of available allocations $\mathcal{C} \subseteq \mathcal{C}_\R$ is a subset of those random allocations which sum to a constant almost surely,\footnote{We want to stress that for all $Y \in \mathcal{C}_\R$ the sum of the stochastic components is deterministic, i.e., for all $Y \in \mathcal{C}_\R,$ there exists $M \in \R$ such that for all  $\omega \in \Omega$ it holds $\sum_{i=1}^N Y_i(\omega) = M.$  We express this fact by simply writing $\sum_{i=1}^N Y_i \in \R$.}
\begin{align*}
    \mathcal{C}_\R = \left\{ Y \in L^0(\R^N) \,\middle|\, \sum_{i=1}^N Y_i \in \R  \right \}.
\end{align*}

Since $\R^N \subseteq \mathcal{C}_\R$, this definition includes the setting of deterministic capital requirements. In general, this notion of systemic risk measures reflects the point of view of a lender of last resort, who would like to reserve some fixed amount of bailout capital to secure the system in the future. However, instead of committing to a fixed allocation today that secures the system under all possible scenarios, the lender can wait which scenario is realised and then distribute the bailout capital. Allowing scenario-dependent allocations can reduce the total capital that needs to be reserved compared to deterministic allocations.

Regarding the aggregation function $\Lambda$, the current literature conceptually distinguishes between two paradigms. One, in which the random variable $X$ represents the risk factors \textit{after} possibly contagious interaction among institutions. Then, the aggregation function is a rather simple function of the risk factors, e.g., a sum, the sum of losses only, a sum of losses where profit is either considered up to some threshold or only a fraction of it, or other similar variations. Examples of such risk measures include 
\cite{brunnermeier2019measuring}, \cite{huang2009framework}, 
\cite{lehar2005measuring}, 
\cite{tarashev2010attributing}.   

On the other hand, there is an approach in which $X$ represents the risk factors \textit{before} any contagious interaction has occurred. Then, the aggregation function is designed to incorporate complex interaction mechanisms that may, for example, include contagion channels like default or illiquidity cascades \cite{amini2016resilience}, \cite{detering2019managing}, \cite{detering2020financial}, \cite{EisNoe}, \cite{frey2018diversification}, \cite{gai2010contagion}, \cite{gai2010liquidity}, \cite{glasserman2015likely}, \cite{hurd2014illiquidity}, \cite{lee2013systemic}, \cite{RogersVeraartGreatesClearingVec}, asset fire sales \cite{braouezec2019strategic}, \cite{caccioli2014stability}, \cite{cifuentes2005liquidity}, \cite{cont2019monitoring}, \cite{cont2016fire}, \cite{detering2021integrated}, \cite{detering2022suffocating}, \cite{feinstein2017financial}, \cite{feinstein2020capital},  or cross-holdings \cite{elsinger2009financial}, \cite{weber2017joint}.

For complex contagion channels, the aggregation function may require additional inputs such as an interbank liability matrix, an asset-bank exposure matrix, or a bank-bank cross-holdings matrix. This is where our first contribution extends the literature. In the setting of random allocations from \cite{biagini2019axiom} and \cite{biagini2020fairness}, we consider an aggregation function that allows for a vector-valued \textit{and} a matrix-valued input, 
\begin{align}\label{line:agg_func_intro}
    \Lambda: \R^N \times \R^{N\times N} \to \R.
\end{align}
We remark that in the context of deterministic allocations such aggregation functions are implicitly covered by the random field structure introduced in \cite{feinstein2017measures}. It allows us to consider aggregation functions that explicitly model contagious interactions in financial networks. Our main focus here is on the Eisenberg--Noe model \cite{EisNoe}, which is the prototype of network-based contagion models.

More specifically, instead of considering vector-valued random risk factors $X \in L^0(\R^N)$, we consider random assets of financial institutions $A \in L^0(\R_+^N)$ together with their explicit interbank lending network given by a random interbank liability matrix $L \in L^0(\R_+^{N \times N})$. Then, we define systemic risk measures as the smallest amount of bailout capital such that random allocations thereof secure the future system,
\begin{align}\label{line:sys_risk_vec_rand_alloc_matrix}
    \rho(A,L) = \inf_{ Y \in \mathcal{C}} \left \{\sum_{i=1}^N Y_i \,\middle|\, \Lambda(A+Y,L)\in \mathbb{A}\right\},
\end{align}
where $\mathcal{C} \subseteq \mathcal{C}_\R$.
Under certain assumptions on $\Lambda$, which in particular include the \textit{total payment shortfall} in the contagion model of \citet{EisNoe}, we show that this systemic risk measure is well-defined
and that there exists a convex set of random allocations $Y \in \mathcal{C}$ that secure the system, i.e., $\Lambda(A+Y,L)\in \mathbb{A},$ while satisfying $\sum_{i=1}^N Y_i = \rho(A,L).$ 
We further reformulate the systemic risk measure in a form that allows to derive an iterative optimisation algorithm for its approximation.

In general, random allocations make such systemic risk measures in (\ref{line:sys_risk_vec_rand_alloc}) challenging to compute numerically. 
Already for deterministic allocations, there exist only few approaches to compute the systemic risk measure. In \cite{feinstein2017measures}, an algorithm is proposed that approximates the efficient frontier of the set-valued risk measure. In \cite{ararat2023computation}, a mixed-integer programming approach is presented to calculate the set-valued risk measure for more general aggregation functions beyond the Eisenberg--Noe model.
There is even less literature concerned with random allocations. In \cite{doldi2023multivariateDeep} and \cite{feng2022deep}, systemic risk measures of the form in (\ref{line:sys_risk_vec_rand_alloc}) are approximated by utilising feedforward neural networks (FNNs) in order to approximate the scenario dependent random allocation as functions $f(X)$ of the risk factors $X$.

We here investigate the extent to which neural networks could be applied to approximate systemic risk measures in our extended setting (\ref{line:sys_risk_vec_rand_alloc_matrix}). Compared to \cite{feng2022deep}, the situation is more complex, since we must rely on neural networks that are able to effectively process the liability matrix as an input. In numerical experiments, it turns out that, despite their universal approximation property, feedforward neural networks are not able to handle this input effectively. However, the structural properties of the problem naturally lead us to consider neural networks, which respect permutation equivariance of the underlying graph data. This motivates the use of graph neural networks (GNNs), an alternative class of neural network architectures that have recently been successfully applied to a wider range of problems within and beyond financial mathematics. For an overview, we refer to the survey papers \cite{wang2021review} and \cite{zhou2020graph}.
Additionally, inspired by \cite{herzig2018mapping}, we derive a structural characterisation of so-called \textit{permutation-equivariant node-labelling functions}. Motivated by this result, we define a permutation-equivariant neural network architecture that exhibits appealing theoretical properties and, in numerical experiments, performs even better than GNNs. We refer to this architecture as (extended) permutation-equivariant neural networks ((X)PENNs).

Before outlining the structure of this article, we summarise our contributions.
\begin{itemize}
    \item We formulate an extension of systemic risk measures with random allocations as in (\ref{line:sys_risk_vec_rand_alloc}), that incorporates more general aggregation functions that allow for random assets and a random interbank liability matrix as inputs, see (\ref{line:sys_risk_vec_rand_alloc_matrix}). In this more realistic setting of random interbank liabilities, we can consider aggregation functions that explicitly model contagion based on the network structure of the institutions.
    In particular, we focus on default contagion as introduced in \citet{EisNoe}.
    Due to the explicitly modelled contagion, our setting emphasises the importance of preventing contagion, instead of paying for its losses afterwards. Additionally, the possibility of random liability matrices opens the door for network reconstruction techniques, especially those that do not provide only one estimator of the \textit{true} network, but many possible ones that fulfil some \textit{boundary conditions}, see for example \cite{cimini2015systemic}, \cite{gandy2017bayesian}, \cite{gandy2019adjustable}, \cite{halaj2013assessing}, or \cite{mastrandrea2014enhanced}. For an overview of network reconstruction techniques see for example \cite{anand2018missing}.
    \item Under certain conditions on the aggregation function (Assumption \ref{assume:aggregate}), which in particular include aggregation functions based on the contagion model of Eisenberg and Noe, such as the \textit{total payment shortfall}, we show that the systemic risk measure is well-defined. 
    In particular, we show existence of ``optimal'' random bailout capital allocations and investigate their properties from a theoretical point of view. We further derive a reformulation of the systemic risk measure that allows us to construct an iterative algorithm for its numerical approximation.
    \item Under mild assumptions, we show that there exists a measurable target function $H^c:\R^N \times \R^{N \times N} \to \R^N$ such that $H^c(A,L)$ is an optimal allocation of the fixed bailout capital $c \in \R.$ Due to its measurability, this allows for approximation via neural networks.
    \item Furthermore, we characterise what we call \textit{permutation-equivariant node-labelling functions} and, motivated by this result, propose the neural network architecture of (X)PENNs. We further prove that (X)PENNs are able to approximate any \textit{permutation-equivariant node-labelling functions}, in particular every measurable function $H^c$, in probability.
    \item Finally, in various numerical experiments we compare the performance of FNNs, GNNs and (X)PENNs and other benchmark approaches in solving different problems related to the approximation of systemic risk measures.
\end{itemize}

This article continues as follows. In Section \ref{section:preliminaries}, we introduce the domain of graphs that is relevant when interpreting financial networks as graphs. Furthermore, we present the contagion model of \citet{EisNoe} and the aggregation functions that we consider in this framework. In Section \ref{section:systemic_risk}, we introduce our systemic risk measures of random financial networks. Section \ref{section:sys_risk_reformulation} provides a reformulation of systemic risk measures that yields an iterative optimisation algorithm for their computation. In Section \ref{section:neural_networks}, we discuss FNNs and GNNs and characterise \textit{permutation-equivariant node-labelling functions}. Motivated by this result we introduce the (X)PENN neural network architecture and discuss its theoretical properties. Section \ref{section:numerical_experiments} contains numerical experiments where we compare the performance of FNNs, GNNs, (X)PENNs and other benchmark approaches. 

%###################################################
\section{Preliminaries}\label{section:preliminaries}
\subsection{Notation}
% \begin{Notes}{Note}
Let $L^p(\R^N) :=L^p(\Omega, \mathcal{F},\mathbb{P}; \mathbb{R}^N)$ denote the space of $\R^N$-valued random variables with finite $p-$norm defined on the probability space $(\Omega, \mathcal{F}, \mathbb{P})$.

For $a,b \in \R^N$, $X,Y \in L^p(\R^N)$ and $c \in \R$ we write for $\triangle \in \{<,>,\leq, \geq, =\}$
\begin{align*}
    a \triangle b &:\iff \forall i\in \{1,\ldots,N\}: a_i \triangle b_i,\\
    X \triangle Y &:\iff \forall i \in \{1,\ldots,N\}: X_i \triangle Y_i \quad \mathbb{P}\text{-a.s.},\\
    a \triangle c &:\iff \forall i\in \{1,\ldots,N\}: a_i \triangle c,\\
    X \triangle c &:\iff \forall i \in \{1,\ldots,N\}: X_i \triangle c \quad \mathbb{P}\text{-a.s.}\\
\end{align*}
For $a,b \in \R^N$, we define $a \,\wedge\,b := \min(a,b)$ and $a \,\vee \, b := \max(a,b)$, where we interpret minimum and maximum componentwise for vectors. 

Furthermore, we define
\begin{align*}
    |a|_p := \left(\sum_{i=1}^N |a_i|^p \right)^{\frac{1}{p}},\quad|X|_p := \left(\sum_{i=1}^N |X_i|^p \right)^{\frac{1}{p}}.
\end{align*}
Similarly, for 
real matrices and matrix valued random variables 
$\ell \in \R^{N \times N'}$ and $L \in L^p(\R^{N \times N'})$, we define
\begin{align*}
    |\ell|_p := \left(\sum_{i=1}^N\sum_{j=1}^{N'} |\ell_{ij}|^p \right)^{\frac{1}{p}},\quad|L|_p := \left(\sum_{i=1}^N\sum_{j=1}^{N'} |L_{ij}|^p \right)^{\frac{1}{p}}.
\end{align*}
Without an index $|\cdot|$ corresponds to $|\cdot|_1.$ In particular $|\cdot|_p$ does not involve taking any expectation for random variables. If we also want to take the expectation, for $X \in L^p$ we define  
$$
\| X\|_p := \myE{}{|X|_p^p}^\frac{1}{p}.
$$
We denote the $i$-th row or column of
$\ell \in \R^{N \times N'}$ and $L \in L^p(\R^{N \times N'})$ as $\ell_{i,\cdot}, L_{i,\cdot}$ and $\ell_{\cdot,i}, L_{\cdot,i}$, respectively.

Furthermore,  
we define $[N] := \{1,\ldots,N\}$ for any $N \in \N$, $\R_+ := \{x \in \R \mid x \geq 0\},$ and $\R_{++}:= \{x\in \R \mid x>0\}.$
% \end{Notes}

\subsection{The domain of graphs}
The financial networks that we consider below
can naturally be described as a graph $g$ containing nodes $1,\ldots,N$ with node features $a_1,\ldots,a_N \in \R^d$ and directed, weighted edges $\ell_{ij} \in \R^{d'}$ between nodes $i,j \in [N]$. Motivated by this, we define the domain of graphs that are directed, weighted and contain node and edge features as follows.
\begin{defn}\label{def:domain_of_graphs}
    We define $\mathcal{D} := \R^{N \times d} \times \R^{N \times N \times d'}$ as the domain of directed, weighted, and featured graphs with $N \in \N$ nodes and with node and edge feature dimensions $d$ and $d'$, respectively. For $g=(a,\ell) \in \mathcal{D}$ we call $a$ the node features and $\ell$ the edge features.
\end{defn}
% \begin{Notes}{Note}
In general $d$ and $d'$ may be multidimensional node and edge features. For example, each node could have as a feature some individual risk coefficient to calculate risk weighted assets, an individual loss given default, or positions in different asset classes. Similarly, each edge could represent not only interbank liabilities, but also cross-holdings, loan specific loss given default or other information. However, in this work we focus on $d=d'=1$, where 
we choose external assets as node features and the liability sizes as edge features.
% \end{Notes}
\begin{rem}
Let $g = (a, \ell) \in \mathcal{D}$. For $i,j \in [N]$ we interpret edges with $\ell_{ij} = (0,\ldots,0) \in \R^{d'}$ as no edge. For applications where differentiation between a zero-edge and no edges is necessary, it would be possible to extend the domain of edge features with some no-edge argument $(a,\ell)\in \R^{N \times d} \times (\R \cup \{\text{NE}\})^{N \times N \times d'}.$ Similarly, we interpret nodes without any edges and with $a_i = (0,\ldots,0)$ as no nodes and hence also account for graphs with less than $N$ nodes. However, if necessary also for nodes a no-node argument NA could be introduced. Obviously, such changes would require all functions operating on the graph domain to be extended accordingly. 
\end{rem}

We are interested in functions on the domain of graphs that assign some value to each node of the network.
\begin{defn}
    We define a node-labelling function $\tau: \mathcal{D} \to \R^{N \times l}, (a,\ell) \mapsto  (\tau_1,\ldots,\tau_N)$ as a function where each component $\tau_n \in \R^l$ corresponds to a label assigned to node $n \in [N].$
\end{defn}
For example, a function that maps each node in a financial network to some node-specific capital requirement would be one example of such a node-labelling function.

As we will see later, an important property of graphs is that, in general, the nodes in a graph do not have any natural order. Imagine we have a financial network with two nodes where one node's assets equal one and the other's amount to two units of capital. From the node with assets 1, there is a liability of 12 units towards the node with assets 2. We can represent this financial network by
\begin{align}\label{line:repr_not_unique_1}
    a = \begin{pmatrix}
        1\\
        2
    \end{pmatrix}, \quad \ell = \begin{pmatrix}
        0 & 12\\ 0& 0
    \end{pmatrix},
\end{align}
where we choose that the node with assets one is the \textit{first} node and the node with assets 2 is the \textit{second} node. However, an equally valid representation of this financial network is
\begin{align}\label{line:repr_not_unique_2}
    a = \begin{pmatrix}
        2\\
        1
    \end{pmatrix}, \quad \ell = \begin{pmatrix}
        0 & 0\\ 12& 0
    \end{pmatrix},
\end{align}
where we altered the node order.

This example showcases that there exist multiple reasonable representations for the same underlying object, when dealing with graph-structured data. Mathematically, we describe this property by introducing permutations.
 
Given any permutation $\sigma \in S_N$, i.e., a bijection from $[N]$ to itself, we denote for any network $g = (a,\ell) \in \mathcal{D}$ the permutation of the network $\sigma(g) = (\sigma(a), \sigma(\ell)) \in \mathcal{D}$ defined by
\begin{align}\label{line:permutation_of_assets}
    &\sigma(a)_{i, k} = a_{\sigma^{-1}(i), k}
\end{align}
for $i \in [N], k \in [d]$ and
\begin{align*}
    \sigma(\ell)_{i,j,k} = \ell_{\sigma^{-1}(i),\sigma^{-1}(j),k},
\end{align*}
for $i,j \in [N], k \in [d'].$

An often desirable property of node-labelling functions is that the value assigned to each node should not depend on the chosen graph representation, but only on the underlying graph.
\begin{defn}\label{def:permutation_equi}
    For any $l \in \N$, we call a node-labelling function $\tau: \mathcal{D}\to\R^{N \times l}$ permutation-equivariant if any permutation $\sigma \in S_N$ of the initial numbering of nodes causes a corresponding permutation of the output,
    \begin{align*}
        \tau(\sigma(g)) = \sigma(\tau(g)), 
    \end{align*}
    where $\sigma(\tau(g)) \in \R^{N \times l}$ is defined analogously to (\ref{line:permutation_of_assets}) by
    \begin{align*}
        &\sigma(\tau(g))_{i, k} = \tau(g)_{\sigma^{-1}(i), k},
    \end{align*}
    for $i \in [N], k \in [l].$
\end{defn}
% \begin{Notes}{Note}
Similar to permutation equivariance, a related important concept in the context of graphs is permutation invariance. It arises naturally, when we deal with functions whose output only depends on the set of inputs, but not on their ordering.
\begin{defn}\label{def:permutation_invari}
    For $d,l \in \N$ we call a function $f: {\R^{N \times d}} \to \R^{l}$ permutation invariant if for any permutation $\sigma \in S_N$ of the inputs
    the result is invariant, i.e.,
    \begin{align*}
        f(\sigma(x)) = f(x),
    \end{align*}
    where $\sigma(x)$ is defined analogously to  (\ref{line:permutation_of_assets}) by
    \begin{align*}
    &\sigma(x)_{i, k} = x_{\sigma^{-1}(i), k}
\end{align*}
for $i \in [N], k \in [d]$.
\end{defn}
Such permutation invariant functions play a role, for example, when we want to map a subset of nodes (a subgraph) to some vector representation or scalar value, that only depends on the set, but not the order of inputs. Examples include the sum, mean, and componentwise min or max operators.\\
% \end{Notes}

A specific subset of nodes that plays an important role for graph neural networks is the so-called neighbourhood of a node. The neighbourhood of a node $i \in [N]$ is defined as the set of all other nodes which have an edge towards $i.$
\begin{defn}
    The neighbourhood of a node $i \in [N]$ in a graph $g = (a, \ell) \in \mathcal{D}$ is defined as
    \begin{align*}
        N_g(i) := \{ j \in [N]\setminus \{i\} | \ell_{ji} \neq 0 \}.
    \end{align*}
\end{defn}
If the referenced graph $g$ is clear in the context, we may simply write $N(i)$ instead of $N_g(i)$.

\subsection{Eisenberg--Noe clearing mechanism}
% \begin{Notes}{Note}
Now, we present the market clearing mechanism from the seminal work of \citet{EisNoe}. It will be crucial in the next section, where we present aggregation functions for financial networks, in particular the \textit{total payment shortfall}. Even though our systemic risk measures are defined in terms of general aggregation functions, our main focus in this paper is on aggregation functions based on the Eisenberg--Noe clearing mechanism.
% \end{Notes}

\begin{defn}[Financial network]
We model a financial network of $N$ institutions and their interbank liabilities as follows:
\begin{enumerate}[label=(\roman*)]
    \item The institutions' assets $a = (a_1,\ldots,a_N) \in \R_+^N$
    \item The interbank liability matrix $\ell \in \R_+^{N\times N},$ where $\ell_{ij} > 0$ indicates a liability of size $\ell_{ij}$ from bank $i$ towards bank $j$. We do not allow self-loops, hence $\diag(\ell) = (\ell_{11},\ldots,\ell_{NN}) = (0,\ldots,0)$.
\end{enumerate}
\end{defn}
Thus, a financial network is represented by a graph $(a,\ell) \in \R_+^N\times \R_+^{N\times N}$ with non-negative node and edge features of dimension one (i.e., $d=d'=1$).
In a network of financial institutions connected by bilateral liabilities, it can occur that assets and incoming cash flows of one participant are not sufficient to meet the full amount of outgoing liabilities. This could lead to a situation where other participants cannot meet their liabilities due to the lack of incoming cash flows, triggering a chain reaction of defaults propagated through the network. To calculate how much value will be actually available for paying off debt at every node in an equilibrium, where everybody pays off as much debt as possible, we can calculate a so-called clearing vector introduced by \citet{EisNoe}. The utilised market-clearing algorithm follows three basic principles:  \textit{limited liability}, which means that an institution can never pay back more than the funds it has available; \textit{proportionality}, which requires that in case of a default all creditors are paid back proportionally; and \textit{absolute priority}, which means that any institution will use all available capital in case its liabilities cannot be fully met.

To describe this clearing vector mathematically, we introduce the following quantities.
For a financial network $(a, \ell) \in \R^N_+ \times \R^{N \times N}_+$,
we define the total liability vector $ \bar \ell \in \R_+^N$ containing the sum of outward liabilities of each node by
\begin{align}\label{line:bar_ell}
        \bar \ell_i &:= \sum_{j=1}^N \ell_{ij}.
\end{align}
Denoting
\begin{align*}
    \mathbf{\Pi}^N := \left\{\pi \in [0,1]^{N \times N} \left |\forall i \in [N]: \pi_{ii}=0, \sum_{j=1}^N \pi_{ij} \in \{0,1\}\right .\right\},
\end{align*}
we define the relative liability matrix $\pi \in \mathbf{\Pi}^N$ by
\begin{align}\label{line:pi}
        \pi_{ij} &:= 
        \begin{cases}
            \ell_{ij}/\bar \ell_i &\text{if $\bar \ell_i > 0$,}\\
            0 \quad & \text{otherwise.}
        \end{cases}
\end{align}
\begin{rem}
    Note that we can always obtain $\ell$ from the pair  $(\pi, \bar \ell)$ and vice versa. Therefore, for every function $f:\R^N_+ \times \R^{N \times N}_+ \to \R^N$, there exists a corresponding function $\bar f: \R^N_+ \times \mathbf{\Pi}^N \times \R^{N}_+ \to \R^N$ such that
    \begin{align*}
        f(a, \ell) = \bar f(a, \pi, \bar \ell).
    \end{align*}
\end{rem}

We define the function $\Phi: \R^N_+  \times \R^N_+ \times \mathbf{\Pi}^N \times \R^N_+ \to \R^N$ by
\begin{align}\label{line:phi}
    \Phi(p, a, \pi, \bar \ell) := \left( {\pi}^Tp + a \right) \wedge  \bar \ell.
\end{align}
This function calculates how much debt could be paid back by each node $i\in [N]$ under the assumption that the other nodes use capital $p_j$, $j\in [N], j \neq i$ to pay off their debt.

The function $\Phi$ is positive, monotone increasing, concave (componentwise, since we map into $\R^N$), and non-expansive (Lipschitz continuous with Lipschitz constant $K=1$) in its arguments $a$ and $\bar \ell.$ Using these properties, in \cite{EisNoe} comparable results for the (unique) fixed point of $\Phi$ are shown.

\begin{defn}
    For given $a \in \R^N_+, \pi \in \mathbf{\Pi}^N, \bar\ell \in \R^N_+$, a fixed point of the function $\Phi(\cdot, a, \pi, \bar\ell)$ is called a clearing vector.
\end{defn}
We can interpret such a clearing vector as a state in which the system is in equilibrium: If each institution uses capital according to the clearing vector to pay back its debts, the fact that a clearing vector is a fixed point of $\Phi$ will ensure that indeed each institution has available capital according to the clearing vector.

The properties of $\Phi$ mentioned above allow \citet{EisNoe} to obtain results regarding the existence of fixed points of this function. In particular, there exist a least and greatest element in the set of fixed points. A clearing vector can be obtained by applying the function iteratively or by solving a finite number of linear equations, see \cite{EisNoe}. Furthermore, they find sufficient conditions for uniqueness of the clearing vector, e.g., $a>0$. 
% in the sense that  $a_i >0$ for every $i \in [N]$.} This is an easy way to ensure unique clearing vectors, h
Hence, from now on we will assume that assets are strictly positive, i.e., $a \in \R_{++}^N$.

The following proposition provides some insights into how a clearing vector depends on the financial network $(a, \pi, \bar\ell).$
\begin{prop}\label{prop:prop_FIX}
Let $[\R^N_+ \to \R_{+}^N]$ denote the space of functions from $\R^N_+ $ to $\R_{+}^N$ and let $\mathcal{P}(\R^N)$ denote the power set of $\R^N.$
    Let further $\operatorname{FIX}: [\R^N_+ \to \R_{+}^N] \to \mathcal{P}(\R^N) $ be the map returning the set of fixed points $\operatorname{FIX}(h)$ of a function $h \in [\R^N_+ \to \R_{+}^N]$.
    Then the following holds.
    \begin{enumerate}[label=(\roman*)]
        \item\label{prop:prop_FIX:in_a} For any $\pi \in \mathbf{\Pi}^N, \bar \ell \in \R^N_+$ the function $f_1:\R^N_{++} \to \R^N$
        \begin{align*}
            f_1(a) = \operatorname{FIX}(\Phi(\cdot, a,\pi,\bar\ell))
        \end{align*}
        is well-defined, positive, monotone increasing, concave and non-expansive.
        \item\label{prop:prop_FIX:in_barell}For any $a \in \R^N_{++}, \pi \in \mathbf{\Pi}^N$ the function $f_2:\R^N_{+} \to \R^N$
        \begin{align*}
            f_2(\bar\ell) = \operatorname{FIX}(\Phi(\cdot, a,\pi,\bar\ell))
        \end{align*}
        is well-defined, positive, monotone increasing, concave and non-expansive.
        \item\label{prop:prop_FIX:in_pi} For any $a \in \R^N_{++}, \bar \ell \in \R^N_+$ the function $f_3: \mathbf{\Pi}^N \to \R^N$ 
        \begin{align*}
            f_3(\pi) = \operatorname{FIX}(\Phi(\cdot, a,\pi,\bar\ell))
        \end{align*}
        is well-defined and continuous.
    \end{enumerate}
\end{prop}
\begin{proof}
    Since for $a \in \R^N_{++}$ the fixed point of $\Phi$ is unique, the well-definedness is clear for all cases. The statements \ref{prop:prop_FIX:in_a} and \ref{prop:prop_FIX:in_barell} follow from Lemma 5 in \cite{EisNoe}.
    Property \ref{prop:prop_FIX:in_pi} follows from Proposition 2.1 in \cite{feinstein2018sensitivity} with the only variation that we define $\mathbf{\Pi}^N$ slightly differently. But since the set $\mathbf{\Pi}^N$ is closed, the same reasoning holds true.
\end{proof}

In the following, we investigate the function that maps a financial network to its unique clearing vector.
\begin{defn}\label{def:CV}
We define the clearing vector functions $\operatorname{CV}: \R^N_{++} \times \R^{N \times N}_+ \to \R^N$ and $\overline{\operatorname{CV}}: \R^N_{++} \times \mathbf{\Pi}^N \times \R^N_+ \to \R^N$ as
\begin{align*}
    \operatorname{CV}(a, \ell) = \overline{\operatorname{CV}}(a, \pi, \bar \ell) = \operatorname{FIX}(\Phi(\cdot, a, \pi, \bar \ell)).
\end{align*}
\end{defn}

From Proposition \ref{prop:prop_FIX} it is clear that $\overline{\operatorname{CV}}$ is continuous with respect to all its components and it is also obvious that $\operatorname{CV}$ is continuous with respect to the assets $a$. 

\begin{rem}\label{rem:pi_row_sum_arbitrary}
    Recall that we chose the relative liability matrix $\pi \in \mathbf{\Pi}^N$ such that for any $i \in [N]$ the $i$-th row $\pi_{i \cdot} \in \R_+^N$ equals 0 or sums to 1, depending on whether $\bar \ell_i$ is 0 or not. However, from the definition of $\Phi$ in (\ref{line:phi}) it is clear that in the case $\bar \ell_i = 0,$ every fixed point $p \in \R_+^N$ of $\Phi(\cdot, a, \pi, \bar \ell)$ must satisfy $p_i=0$ for any choice of $\pi_{i\cdot} \geq 0$. Subsequently $\overline{\operatorname{CV}}(a, \pi, \bar \ell)$ is the same for any choice of $\pi_{i\cdot} \geq 0$ if $\bar \ell_i = 0$. 
    % This observation will help us prove the following proposition, where we show that  $\operatorname{CV}$ is continuous with respect to the liability matrix $\ell.$ 
\end{rem}
\subsection{Aggregation functions}
One crucial element in the construction of systemic risk measures (\ref{line:sys_risk_vec_rand_alloc_matrix}) is the aggregation rule (\ref{line:agg_func_intro}). In the following, we first specify the general family of aggregation functions that we consider in this paper before we state some important examples based on the Eisenberg--Noe clearing mechanism. To ensure that we have unique clearing vectors for those Eisenberg--Noe-based examples, we consider only aggregation functions for financial networks with strictly positive assets, i.e., $a \in \R_{++}^N.$

\begin{assume}\label{assume:aggregate}
    For the aggregation function $\Lambda:\R_{++}^N\times\R_+^{N\times N} \xrightarrow{}\R$, we make the following assumptions.
    \begin{enumerate}[label=(\roman*)]
        \item $\Lambda$ is convex and decreasing in its first component. \label{assume:aggregate:convex/incre}
        % \item $\Lambda$ is continuous. \label{assume:aggregate:cont}
        \item For every $u < \infty$, we can find a deterministic allocation $y_u^* \in \R_+^N$ such that $\Lambda(\cdot + y_u^*,\ell) = 0$ for every $\ell \in \R_+^{N\times N}$ with $\ell \leq u$. \label{assume:aggregate:finite_zero}
        % $\forall i,j \in [N]: \ell_{i,j} \leq u.$ 
        \item There exist coefficients $c_1, c_2\geq 0$ such that $\Lambda(\cdot,\ell) \leq c_1|\ell|_1 + c_2.$ \label{assume:aggregate:growth}
    \end{enumerate}
\end{assume}
The interpretation of convexity is clear.
The other two properties are similar and describe how the aggregation function depends on the liability matrix. This will be important in the subsequent chapters, when we consider random networks $(A, L)$ instead of the deterministic networks $(a, \ell).$ 
Property \ref{assume:aggregate:finite_zero} can be explained as follows:
If we know that all interbank connections are at most of size $u>0$, then we know how much debt each node can be in at most and hence we can always set the aggregated loss to 0 by providing enough money to each node in the network. The linear-growth property \ref{assume:aggregate:growth} guarantees that the aggregated loss is smaller than an affine function of all interbank liabilities, which makes sense if we assume that every loss originates from unpaid debt. It guarantees that, for a random liability matrix $L \in L^1$, also $\Lambda(A,L) \in L^1$, for any random asset vector $A$.\\

One particularly interesting family of aggregation functions, that includes an explicit modelling of the default contagion, is based on the Eisenberg--Noe clearing mechanism. In the following, we present some examples of such aggregation functions.
\begin{exa}\label{exa:lambda:totalpaymentshortfall}
    The \textit{total payment shortfall}, which we will use in our numerical case studies, is calculated by first computing the clearing vector $\operatorname{CV}(a, \ell)$ of the network and then subtracting it from the total liability vector $\bar \ell$. This non-negative vector represents, in each component, the amount of liabilities towards other nodes that cannot be paid for. If we sum up these values, we obtain the \textit{total payment shortfall} of the network.
    Let $\operatorname{CV}: \R_{++}^N\times\R_+^{N\times N} \xrightarrow{}\R^N$ be the function returning the clearing vector. Then, the \textit{total payment shortfall} $\Lambda: \R_{++}^N\times\R_+^{N\times N} \xrightarrow{}\R$ is defined as
\begin{align*}
     \Lambda(a, \ell) = \sum_{i=1}^N \bar \ell_i - \operatorname{CV}_i(a, \ell),
\end{align*}
where $\bar \ell\in \R^N$ is the total liability vector associated with $\ell \in \R^{N \times N}_+$, see (\ref{line:bar_ell}).

Property \ref{assume:aggregate:convex/incre} of Assumption \ref{assume:aggregate} is trivially fulfilled by the \textit{total payment shortfall}, since it follows directly from the concavity of $\operatorname{CV}$. 
Property \ref{assume:aggregate:finite_zero} holds because, if we know that $\bar \ell_i = \sum_{j=1}^N \ell_{ij} \leq Nu$, then for any non-negative vector $p \in \R_+^N$ and for $y \in \R_+^N$ with $y \geq Nu$, it holds that
     \begin{align*}
          \Phi(p, a + y, \ell)  &= \Big( \pi^Tp + a+y \Big) \wedge  \bar \ell
          = \Big( 
         \underbrace{\pi^Tp}_{\geq 0} + 
         \underbrace{a}_{>0}+
         \underbrace{\big(y - \bar \ell}_{\geq 0}\big) + 
         \bar \ell  \Big) \wedge  \bar \ell
         = \bar \ell.
     \end{align*}

    Consequently, the only possible fixed point and hence value of $\operatorname{CV}$ is $\operatorname{CV}(a+y, \ell)= \bar \ell$,
    and we get
    \begin{align*}
        \Lambda(a + y,\ell) = \sum_{i=1}^N \bar\ell_i - \operatorname{CV}_i(a+y ,\ell) = 0.
    \end{align*}
 Finally, Property \ref{assume:aggregate:growth} holds because $\Lambda(a,\ell) = \sum_{i=1}^N \bar \ell_i - \operatorname{CV}_i(a, \ell) \leq |\ell|_1$. 
\end{exa}
\begin{exa}\label{exa:lambda:2}
    Another possible choice for the aggregation function $\Lambda$ is to interpret one special node $j^* \in [N]$ in the system as the society, and to consider only the shortfall to this society, as proposed in \cite{ararat2020dual}. Then with the relative liability matrix $\pi \in \mathbf{\Pi}^{N} $ from Equation (\ref{line:pi}),
    $$
    \Lambda(a,\ell) = \sum_{i \neq j^*} \pi_{i,j^*}\left( \bar\ell_i - \operatorname{CV}_i(a,\ell)\right).
    $$
    For this example, Properties \ref{assume:aggregate:convex/incre}, and \ref{assume:aggregate:finite_zero} of Assumption \ref{assume:aggregate} hold, with the same reasoning as in Example \ref{exa:lambda:totalpaymentshortfall}. Property \ref{assume:aggregate:growth} holds, because 
    $$
    \sum_{i \neq j^*} \underbrace{\pi_{i,j^*}}_{\leq1}\left( \bar\ell_i - \operatorname{CV}_i(a,\ell)\right) \leq |\ell|_1.
    $$
\end{exa}
\begin{exa}\label{exa:lambda:3}
    More generally, each node's payment shortfall and the summed shortfalls could be scaled with some individual or systemic risk coefficients $c_1,\ldots,c_N,c \in \R_+,$
    $$
    \Lambda(a,\ell) = c\sum_{j=1}^N c_j\sum_{i=1}^N \pi_{i,j}\left( \bar\ell_i - \operatorname{CV}_i(a,\ell)\right).
    $$
    In this example, again, Properties \ref{assume:aggregate:convex/incre}, and \ref{assume:aggregate:finite_zero} of Assumption \ref{assume:aggregate} hold with the same reasoning as in Example \ref{exa:lambda:totalpaymentshortfall}. Property \ref{assume:aggregate:growth} holds, because
    $$
    c\sum_{j=1}^N c_j\sum_{i=1}^N \pi_{i,j}\left( \bar\ell_i - \operatorname{CV}_i(a,\ell)\right) \leq \max\{c,c_1,\ldots,c_N\}^2|\ell|_1.
    $$
\end{exa}
\begin{rem}
    Other interesting aggregation functions for which, however, Property \ref{assume:aggregate:growth} is not necessarily true anymore are obtained by applying general risk-averse functions to each node's shortfall  
    $$
    \Lambda(a,\ell) = \sum_{j=1}^N u_j\left(\sum_{i=1}^N \pi_{i,j}\left( \bar\ell_i - \operatorname{CV}_i(a,\ell)\right)\right),
    $$
    to the \textit{total payment shortfall}
    $$
    \Lambda(a,\ell) = u\left(\sum_{i=1}^N \bar\ell_i - \operatorname{CV}_i(a,\ell)\right),
    $$
    or to the loss caused by each node
    $$
    \Lambda(a,\ell) =  \sum_{i=1}^N u_i\left( \bar\ell_i - \operatorname{CV}_i(a,\ell)\right).
    $$
    Here $u,u_1,\ldots,u_N: \R \to \R$ are increasing and convex functions. Popular choices would, for example, be exponential functions of the form
    $$
    u(x)= e^{\gamma x},
    $$
    with risk-aversion coefficients $\gamma,\gamma_1,\ldots,\gamma_N \geq 0.$ 

    This aggregation functions would also fulfil Properties \ref{assume:aggregate:convex/incre}, and \ref{assume:aggregate:finite_zero} of Assumption \ref{assume:aggregate} with the same reasoning as in Example \ref{exa:lambda:totalpaymentshortfall}. However, depending on the behaviour of the functions $u,u_1,\ldots,u_N$, Property \ref{assume:aggregate:growth} might not hold. It would still be possible to recover all theory in the subsequent sections without Property \ref{assume:aggregate:growth} if we assume that the liability matrix $L$ is bounded, i.e., $L \in L^\infty$ instead of only $L \in L^1.$ For more information on the $L^\infty$ setting, we refer to Remark \ref{rem:L^1vsL^infty}.
\end{rem}
% \end{Notes}

%###################################################
\section{Systemic risk of random financial networks}\label{section:systemic_risk}
In this section, we specify and analyse the systemic risk measures on random financial networks of type (\ref{line:sys_risk_vec_rand_alloc_matrix}) that we consider in this paper.
\begin{defn}
We model a random network $G=(A,L)$ of $N\in \N$ financial institutions as
\begin{enumerate}[label=(\roman*)]
    \item the institutions' assets $A = (A_1,\ldots,A_N) \in L^0(\R_{++}^N)$; and
    \item the interbank liability matrix $L \in L^0(\R_+^{N\times N})$, where $L_{ij}$ denotes the size of the liability 
    from bank $i$ to bank $j$. 
    We do not allow self loops, hence $\diag(L) = (L_{11},\ldots,L_{NN}) = (0,\ldots,0)$.
\end{enumerate}
\end{defn}
% \begin{Notes}{Note}
For the remainder of the paper, we make the following assumptions. 
\begin{assume}\label{assume}\
    \begin{enumerate}[label=(\roman*)]
    \item \label{assume:eta}  $\eta : L^1(\Omega, \mathcal{F}, \mathbb{P}; \R^N) \to \R$ is a univariate risk measure that is
    \begin{itemize}
        \item monotone increasing: for $X>Y: \eta(X) > \eta(Y)$;
        \item convex: for $\lambda \in [0,1]:$ $\eta(\lambda X + (1-\lambda)Y) \leq \lambda\eta(X) + (1-\lambda)\eta(Y)$;
        \item normalised: $\eta(0) = 0$.
    \end{itemize}
    We define the acceptance set as
    \begin{align*}
        \mathbb{A} := \{ Z \in L^1(\Omega, \mathcal{F},\mathbb{P}; \R) \mid \eta(Z) \leq b\},
    \end{align*}
    for some acceptable risk threshold $b > 0$.
    \item The liability matrix is integrable, i.e., $L \in L^1(\R_+^{N \times N})$.
    We denote the set of all random financial networks with integrable liability matrix by
        \begin{align}\label{def:set_of_graphs}
            \mathcal{G} := \{(A,L) \mid
            A \in L^0(\R_{++}^N)
            ,L \in L^1(\R_+^{N\times N})
            ,\diag(L) = 0\},
        \end{align}
    and define the random vector of total liabilities $(\bar L_1,\ldots,\bar L_N)^T$ componentwise for $i \in [N]$ as
\begin{align*}
    \bar L_i := \sum_{j=1}^N L_{ij}.
\end{align*}
 \item The aggregation function $\Lambda$ fulfils all properties in  Assumption \ref{assume:aggregate}.
 \item The set of admissible allocations $\mathcal{C}$ consists of all non-negative allocations $Y$ whose stochastic components sum to some fixed value almost surely:
    \begin{align*}
         \mathcal{C} := \left\{ Y \in L^0(\R_+^N) \,\left|\, \sum_{n=1}^N Y_n \in \R \right. \right \} \subseteq \mathcal{C}_\R.
    \end{align*}
\end{enumerate}
\end{assume}
There are two reasons why we only consider non-negative allocations in $\mathcal{C}$. On one hand, cross-subsidisation -- where one bank pays for the losses of another bank -- is unrealistic behaviour. On the other hand, the aggregation function -- in particular, the clearing vector -- is only defined for networks with non-negative assets. By restricting $Y\geq0$, we ensure that $A+Y\geq0$.

Then, under these assumptions, systemic risk measures of random financial networks introduced in (\ref{line:sys_risk_vec_rand_alloc_matrix}) are defined as follows. 
\begin{defn}\label{def:sys_risk_DEF}
 The systemic risk measure $\rho_b: \mathcal{G} \to
 \R \cup \{\pm \infty\}$ on random financial networks $\mathcal{G}$ as in (\ref{def:set_of_graphs}) 
 is defined by
\begin{align}\label{def:sys_risk_b}
    \rho_b(G) = \rho_b(A,L) := \inf_{ Y \in \mathcal{C}} \left \{\left.\sum_{n=1}^N Y_n \right| \eta \left({\Lambda(A+Y,L)}\right) \leq b  \right\}, \quad G=(A,L) \in \mathcal{G},
\end{align}
where $\eta, \Lambda$, and  $\mathcal{C}$ fulfil Assumption \ref{assume}, and $b>0$.
\end{defn}

\begin{rem}\label{rem:continuity_eta}
    From the extended Namioka--Klee Theorem (Theorem 1 in \cite{biagini2010extension}), we know that any $\eta: L^p \to \R$ with the properties in Assumption \ref{assume} \ref{assume:eta} is continuous in $L^p$ for  $1\leq p \leq \infty$, in particular for $p=1.$ This means that
    \begin{align*}
        \forall \varepsilon > 0 \, \exists \delta > 0: \|X-Y\|_1<\delta \implies |\eta(X)-\eta(Y)|< \varepsilon.
    \end{align*}
    Note that one additional popular property in Assumption \ref{assume} \ref{assume:eta} would be cash invariance, defining the class of monetary risk measures.
    This property is often assumed in the risk measure literature, but is not necessary to derive the subsequent results.
\end{rem}
\begin{rem}\label{rem:L^1vsL^infty}
    We present systemic risk measures in the $L^1$ framework. However, it is also possible to define $\eta$ on $L^\infty$ and to require $L \in L^\infty$ instead of $L \in L^1.$ In this simplified setting, all results still hold. In particular, some assumptions can be relaxed. For example, it is possible to choose $b\geq0$ instead of $b>0$ in (\ref{def:sys_risk_b}), and we can drop Property \ref{assume:aggregate:growth} in Assumption \ref{assume:aggregate}. 
\end{rem}
% \end{Notes}

For a deterministic liability matrix $\ell \in \R_+^{N \times N}$, Definition \ref{def:sys_risk_DEF} aligns well with existing definitions of systemic risk measures for vector-valued risk factors; see for example (\ref{line:sys_risk_vec_rand_alloc}). In particular, following \cite{biagini2019axiom}, we obtain the following result.
\begin{prop}\label{prop:embedding}
    For a deterministic liability matrix $\ell\in \R_+^{N \times N}$ the systemic risk measure 
    \begin{align*}
        \rho:L^0(\R_{++}^N) \to \R \cup \{\pm \infty\}\\
        \rho(A) :=  \rho_b(A,\ell)
    \end{align*}
    is monotone and convex on $\{\rho(A)< +\infty \}$, i.e., $\rho$ is a convex systemic risk measure.
\end{prop}
\begin{proof}[Proof of Proposition \ref{prop:embedding}]
    We can apply Lemma 3.3 from \cite{biagini2019axiom} and only need to show the following two properties:
    \begin{itemize}
        \item[(P1)] For all $Y \in \mathcal{C}$ and the set $$\mathcal{A}^Y = \{A\mid A \in  L^0(\R_+^N), \eta\left(\Lambda(A+Y,\ell)\right)\leq b\}$$ it holds: $A_2 \geq A_1 \in \mathcal{A}^Y \implies A_2 \in \mathcal{A}^Y.$
        \item[(P2)] For all $Y^1,Y^2 \in \mathcal{C}$, all $A^1,A^2 \in  L^0(\R_+^N)$ with $\eta\left(\Lambda(A^i+Y^i,\ell)\right)\leq b, i=1,2$, and all $\lambda \in [0,1]$ there exists $Y \in \mathcal{C}$ such that $|Y|_1 \leq \lambda |Y^1|_1 + (1-\lambda)|Y^2|_1$ and $\eta\left(\Lambda(\lambda A^1 + (1-\lambda)A^2 + \lambda Y^1 + (1-\lambda)Y^2,\ell)\right)\leq b$.
    \end{itemize}
    Since $\Lambda$ is monotone decreasing for $A^1 \leq A^2$, it follows immediately that $\eta\left(\Lambda(A^2+Y,\ell)\right)\leq \eta\left(\Lambda(A^1+Y,\ell)\right)$, which shows (P1).
    For (P2) we choose $Y = \lambda Y^1 + (1-\lambda)Y^2$. Then, since $\Lambda$ is convex and $\eta$ is monotone increasing and convex, we get
    \begin{align*}
        &\eta\left(\Lambda(\lambda A^1 + (1-\lambda)A^2  + \lambda Y^1 + (1-\lambda)Y^2,\ell)\right)\\
        &\leq \eta\left(\Lambda(\lambda (A^1 + Y^1) + (1-\lambda)(A^2   + Y^2),\ell)\right)\\
        &\leq \lambda \eta\left(\Lambda(A^1 + Y^1 ,\ell)\right) + (1-\lambda)\eta\left(\Lambda(A^2   + Y^2,\ell)\right) \leq b,
    \end{align*}
    which shows (P2).
\end{proof}
It may seem restrictive that the previous result only holds on $\{\rho(X)< + \infty \}$, but in fact, Proposition \ref{prop:risk_is_well_def} shows that $\rho_b(A,L) < +\infty$ for all $(A,L) \in \mathcal{G}$.

Let us now return to the setting where the liability matrix is random. For $G \in \mathcal{G}$, the systemic risk $\rho_b(G)$ is well-defined in the following sense.
\begin{prop}\label{prop:risk_is_well_def}
    Let $G \in \mathcal{G}$ and $b > 0.$
    Then, $-\infty < \rho_b(G) < +\infty.$
\end{prop}
% \begin{Notes}{Note}
\begin{proof} 
    By definition, it is clear that $\rho_b(G)\geq 0> - \infty.$
    We show that $\rho_b(G)<+ \infty.$

    We know from Property \ref{assume:aggregate:growth} of Assumption \ref{assume:aggregate} that there exist coefficients $c_1, c_2\geq 0$ such that $\Lambda(\cdot,L) \leq c_1|L|_1 + c_2$ which, in particular, means that $\Lambda(\cdot,L) \in L^1$.
    By monotone convergence, it is clear that, for $M \in \R$,
    $$
    \| (c_1 |L|_1+c_2) \mathds{1}_{\{L \nleq M\}} \|_1 = \| (c_1 |L|_1+c_2) (1-\mathds{1}_{\{L \leq M\}}) \|_1 
    \xrightarrow[]{M \rightarrow \infty} 0. 
    $$
    Therefore, by continuity of $\eta$ (see Remark \ref{rem:continuity_eta}), we can choose $M^*>0$ such that
    $$
    \eta \left((c_1 |L|_1 + c_2) \mathds{1}_{\{L \nleq M^*\}} \right) < b.
    $$
    Additionally, since $L\mathds{1}_{\{L \leq M\}} \leq M$, we know from Property \ref{assume:aggregate:finite_zero} of Assumption \ref{assume:aggregate} that there exists a deterministic allocation $y_{M^*} \in \R^N_+$ such that
    $$
        \Lambda(A+y_{M^*},L)\mathds{1}_{\{L \leq {M^*}\}} = 0.
    $$
    Together, since $\eta$ is increasing, this means that
    \begin{align*}
        \eta\left(\Lambda(A +y_{M^*},L)\right) &=  \eta\left(\Lambda(A +y_{M^*},L) \mathds{1}_{\{L \nleq {M^*}\}} + \underbrace{\Lambda(A +y_{M^*},L) \mathds{1}_{\{L \leq {M^*}\}}}_{=0}\right)\\
        &\leq \eta \left((c_1 |L|_1 + c_2) \mathds{1}_{\{L \nleq {M^*}\}}\right) 
        < b,
    \end{align*}    
    Hence, $y_{M^*}$ secures the network and, in particular, 
    $$
    \rho_b(G) \leq |y_{M^*}|_1  < +\infty,
    $$
    which completes the proof.
\end{proof}  
% \end{Notes}
The following corollary shows that, in order to compute the systemic risk, we can restrict the set of random allocations to those bounded by some value $u \in \R_+$, provided it is sufficiently large.
\begin{cor}\label{corr:can_bound_allocations}
    Let $G=(A,L) \in \mathcal{G}$, $b > 0$, and let $\rho_b(G)<u<+\infty$.
    Then
    \begin{align*}
     \rho_b(G) &= \inf \left \{\left.\sum_{n=1}^N Y_n \,\right|\,Y \in \mathcal{C}, \eta\left(\Lambda(A+Y,L)\right) \leq b  \right\}\\
     &= \inf \left \{\left.\sum_{n=1}^N Y_n \,\right|\,Y \in \mathcal{C}, Y \leq u, \eta\left(\Lambda(A+Y,L)\right) \leq b  \right\}.
\end{align*}
\end{cor}
\begin{proof}
    It is clear that 
    \begin{align*}
     &\inf \left \{\left.\sum_{n=1}^N Y_n \,\right|\,Y \in \mathcal{C}, \eta\left(\Lambda(A+Y,L)\right) \leq b  \right\}\\
     \leq &\inf \left \{\left.\sum_{n=1}^N Y_n \,\right|\,Y \in \mathcal{C}, Y \leq u, \eta\left(\Lambda(A+Y,L)\right) \leq b  \right\}.
\end{align*}

    To see the reverse inequality, let $(Y^{(k)})_{k \in \N} \subseteq \mathcal{C}$ be such that $\eta\left(\Lambda(A+Y^{(k)},L)\right) \leq b$ for all $k \in \N$,
    and $|Y^{(k)}|_1 \xrightarrow[]{k \rightarrow \infty} \rho_b(G)$.
    
    Without loss of generality, we can impose that $(Y^{(k)})_{k \in \N}$  converges monotonically from above.
    Then, there exists some $k^* \in \N$ such that $|Y^{(k)}|_1 \leq u$ for all $k \geq k^*$.
    Define 
    $$
    \tilde Y^{(k)} = \begin{cases}
        Y^{(k)},\quad k \geq k^*\\
        Y^{(k^*)}, \quad k < k^*.
    \end{cases}
    $$
    The sequence $(\tilde Y^{(k)})_{k \in \N}$ is bounded by $u$, because a vector of non-negative entries summing up to a value $u$ is trivially bounded by $u$ in every component. Furthermore,  $(|\tilde Y^{(k)}|_1)_{k \in \N}$ converges to $\rho_b(G)$ as well, and hence
    \begin{align*}
     &\inf \left \{\left.\sum_{n=1}^N Y_n \,\right|\,Y \in \mathcal{C}, \eta\left(\Lambda(A+Y,L)\right) \leq b  \right\}\\
     \geq &\inf \left \{\left.\sum_{n=1}^N Y_n \,\right|\,Y \in \mathcal{C}, Y \leq u, \eta\left(\Lambda(A+Y,L)\right) \leq b  \right\},
\end{align*}
    which concludes the proof.
\end{proof}

Now, we want to discuss some further properties of the systemic risk measure as defined in (\ref{def:sys_risk_b}). In particular, we will prove the existence of an optimal random allocation for the systemic risk $\rho_b(G)$, and therefore, the infimum in (\ref{def:sys_risk_b}) is actually a minimum. To this end, we define optimal allocations as follows.
\begin{defn}
    We call an allocation $Y$ optimal for $G \in \mathcal{G}$ with respect to the systemic risk measure defined in (\ref{def:sys_risk_b}) if and only if $Y \in \mathcal{C}$, $|Y|_1 = \rho_b(G)$ and $\eta\left(\Lambda(A+Y,L)\right)\leq b$.
\end{defn}
Furthermore, while it is not necessary that an optimal allocation should be unique, we can show that the set of optimal allocations is convex, and that, for an optimal allocation $Y$, the risk of loss $\eta\left(\Lambda(A+Y,L)\right)$ is exactly $b$. The following proposition summarises these results.
\begin{prop}\label{prop:min}
Let $G=(A,L) \in \mathcal{G}$, $b > 0$, and $\rho_b(G)$ be the systemic risk measure from (\ref{def:sys_risk_b}). Then the following holds:
    \begin{enumerate}[label=(\roman*)]
        \item There exists an optimal $Y^*$ such that the infimum $\rho_b(G)$ is attained, i.e., it is a minimum.
        \item For each optimal allocation $Y^*$ the risk of loss is exactly $b$:
        \begin{align*}
           \eta\left(\Lambda(A+Y^*,L)\right) = b.
        \end{align*}
        \item The set of all optimal allocations is a convex set.
    \end{enumerate}
\end{prop}

\begin{proof}
    Let $G=(A,L) \in \mathcal{G}$, $b > 0$, and $Y^{(k)} \subset \mathcal{C}$ be a sequence such that
    $\forall k \in \N: \eta\left(\Lambda(A+Y^{(k)}, L)\right) \leq b,$ and $|Y|_1^{(k)} \xrightarrow[k\rightarrow \infty]{} \rho_b(G).$
    
    Without loss of generality, we choose the sequence such that $|Y|_1^{(k)}$ is monotonically decreasing towards $\rho_b(G)$.

    By Corollary \ref{corr:can_bound_allocations}, we can choose these allocations uniformly bounded by some $u \in \R$, $u>\rho_b(G)$. Then, since $0 \leq Y^{(k)} \leq u$ is bounded, according to Koml{\'o}s' Theorem \ref{komlos_thm} we can find a new sequence 
    \begin{align}\label{line:use_komlos}
        \tilde Y^{(k)} \in \operatorname{conv}\{Y^{(k)}, Y^{(k+1)},\ldots\}
    \end{align}
    that converges almost surely to a limit random variable $ \tilde Y^*$, with $\operatorname{conv}$ being the set of convex combinations define in (\ref{def:conv}).

    It is clear that $\mathcal{C}$ is convex, hence we know that $\forall k: \tilde Y^{(k)} \in \mathcal{C}$. Furthermore, since $\eta$ is convex and $\Lambda(A+Y,L)$ is convex in $Y$, we get that, for $n_k \in \N$, $j\in \{1,\ldots,n_k\}$, $k_j \in \{k,k+1,\ldots\}$, $a_{k_j} \geq 0$, $\sum_{j=1}^{n_k}a_{k_j}=1$, such that
    \begin{align*}
        \tilde Y^{(k)} = \sum_{j=1}^{n_k} a_{k_j} Y^{(k_j)},
    \end{align*}
    it holds that
    \begin{align*}
       \eta\left(\Lambda(A+\tilde Y^{(k)},L)\right) &= \eta\left(\Lambda(A+ \sum_{j=1}^{n_k} a_{k_j} Y^{(k_j)},L)\right)\\
        &\leq\eta\left(\sum_{j=1}^{n_k} a_{k_j}\Lambda\left(A+ Y^{(k_j)},L\right)\right) \\
        % &\leq \eta\left(\sum_{i=1}^N \bar L_i- \sum_{j=1}^{n_k} a_{k_j}\operatorname{CV}_i\left(A+ Y^{(k_j)},L\right)\right) \\
        &\leq \sum_{j=1}^{n_k} a_{k_j} \eta\left(\Lambda(A+\tilde Y^{(k_j)},L)\right) \leq b.
    \end{align*}
    
    Since we choose $(Y^{(k)})_k$ such that $\sum_{i=1}^N Y^{(k)}_i$ is decreasing from above to $\rho_b(G)$, we know that
    \begin{align*}
        &\rho_b(G) \leq \sum_{i=1}^N \tilde Y_i^{(k)} =  \sum_{i=1}^N \sum_{j=1}^{n_k} a_{k_j} Y_i^{(k_j)} = \sum_{j=1}^{n_k} a_{k_j}  \sum_{i=1}^N Y_i^{(k_j)} \\
        &\leq \sum_{i=1}^N Y_i^{(k)} \xrightarrow[k\rightarrow \infty]{} \rho_b(G).
    \end{align*}
 % Hence, the sequence of $\tilde Y^{(k)}$ fulfils the same requirements as $Y^{(k)}$ and 
 For the rest of the proof we will only consider the converging sequence of $\tilde Y^{(k)}$ with limit $ \tilde Y^*$, but to simplify the notation we write $Y^{(k)}$ and $Y^*$ in place of $\tilde Y^{(k)} $ and $\tilde Y^*$.
\begin{enumerate}[wide, label=(\roman*)] 
    \item First, it is clear that $Y^*$ is in $\mathcal{C}$ and bounded, because $\mathcal{C}$ is clearly closed and because we know that, for all $k\in \N$, $0 \leq Y^{k} \leq u$, and hence $u \geq Y^* \in L^\infty(\R_+^N)$. Furthermore, by construction, the components of $Y^*$ sum to the constant value of $\rho_b(G)$,
    \begin{align*}
        |Y^*|_1 = \sum_{i=1}^N Y_i^{\infty} =  \sum_{i=1}^N \lim_{k\rightarrow \infty} Y_i^{(k)} = \lim_{k\rightarrow \infty} \sum_{i=1}^N Y_i^{(k)} = \rho_b(G).
    \end{align*}
    Next, we show that $Y^*$ actually fulfils the risk constraint $\eta\left(\Lambda(A+Y^*, L)\right) \leq b.$
    We assume $\eta\left(\Lambda(A+Y^*, L)\right) = b + \varepsilon$ for some $\varepsilon > 0$ and lead this to a contradiction. 
    
    For any $k \in \N$, we have
    $$
    \eta\left(\Lambda(A+Y^*,L)\right) - \eta\left(\Lambda(A+ Y^{(k)},L)\right) \geq \varepsilon.
    $$

Furthermore, since
$
Y^{(k)} \xrightarrow[]{k \to \infty} Y^*
$
and since $\Lambda(A+ \cdot,L)$ is continuous, it also holds that
$$
\Lambda(A+Y^{(k)},L) \xrightarrow[]{k \to \infty} \Lambda(A+Y^*,L).
$$
Using the fact that $\Lambda(\cdot,L) \leq c_1|L|_1 +c_2 \in L^1(\R_+^{N \times N})$, we apply the dominated convergence theorem to obtain
$$
\|\Lambda(A+Y^{(k)},L) -  \Lambda(A+Y^*,L)\|_1 \xrightarrow[]{k \to \infty} 0.
$$
But, by the continuity of $\eta$ (Remark \ref{rem:continuity_eta}), this would mean that
$$
\eta\left(\Lambda(A+Y^{(k)},L)\right) \xrightarrow[]{k \to \infty} \eta\left(\Lambda(A+Y^*,L)\right)
$$
which contradicts 
$$
\forall k \in \N: \eta\left(\Lambda(A+Y^*,L)\right) - \eta\left(\Lambda(A+ Y^{(k)},L)\right) \geq \varepsilon.
$$
    Therefore it must hold that $\eta\left(\Lambda(A+Y^*,L)\right) \leq b.$

    This concludes the first part of the proof, where we showed that the infimum is actually a minimum. 

    \item Now, we continue and show that the risk actually equals $b$. We already showed ``$\leq$''. In order to show ``$\geq$'', assume $\eta\left(\Lambda(A+Y^*,L)\right)= b - \varepsilon$. 
    By continuity of $\eta$ (Remark \ref{rem:continuity_eta}), we know for $\varepsilon/2$, there exists a $\delta>0$ such that for any $\widetilde Y \in \mathcal{C}$
    $$
    \left\|\Lambda(A+Y^*,L) - \Lambda(A+\widetilde{Y},L)\right\|_1 < \delta \implies \left|\eta\left(\Lambda(A+Y^*,L)\right) - \eta(\Lambda(A+\widetilde{Y},L))\right| < \varepsilon/2.
    $$
   
    Furthermore, by the continuity of $\Lambda(A + \cdot,L)$ we can find $\delta'$ such that
    $$
        \left|\Lambda(A+Y^*,L) - \Lambda(A+\widetilde{Y},L)\right| \leq \delta/2
    $$
    for all $Y^*, \widetilde{Y}$ with 
    $$
        \left| Y^* - \widetilde{Y} \right|_1 < \delta',
    $$
    which is, in particular, true for
    \begin{align*}
        \widetilde{Y} = Y^* - \begin{pmatrix}
            \delta'/2\\0\\ \vdots \\ 0
        \end{pmatrix} =Y^*  - \frac{\delta'}{2} e_1,
    \end{align*}
    where $e_1 = (1,0,\ldots,0)^T \in \R^N.$
    Therefore, by the continuity and monotonicity of $\eta$ it holds that
    $$
    \eta\left(\Lambda(A+\widetilde{Y},L)\right) \leq b - \varepsilon + \varepsilon/2 \leq b.
    $$
    However, this means that $\widetilde{Y}$ is a cheaper bailout strategy than $Y^*$ (it sums to $\rho_b(G) - \delta'/2$ instead of $\rho_b(G)$) which makes the system acceptable. But then $\rho_b(G)$ cannot be the infimum, which is a contradiction. Therefore, it must be true that $\eta\left(\Lambda(A+Y^*,L)\right) \geq b$, and hence $\eta\left(\Lambda(A+Y^*,L)\right)= b.$
    
    \item In the last part of the proof, we show that the set of optimal allocations is convex.
    Let $Y^{(1)}, Y^{(2)}$ be two optimal allocations for $G \in \mathcal{G}$ with respect to the systemic risk measure $\rho_b(G)$. 
    Let $\lambda \in (0,1)$ and $\widetilde{Y} = \lambda Y^{(1)} + (1-\lambda) Y^{(2)}.$ 
        Thus, we obtain        
        \begin{align*}
            &\eta\left(\Lambda(A+\widetilde{Y},L)\right) = \eta\left(\Lambda\left(\lambda (A+Y^{(1)}) + (1-\lambda)(A+Y^{(2)}),L\right)\right)\\
            &\leq \eta\left(
            \lambda \Lambda\left(A+Y^{(1)},L \right) + (1-\lambda)\Lambda\left(A+Y^{(2)},L \right)
            \right)\\
            &\leq \lambda \eta\left(\Lambda(A+Y^{(1)},L)\right) + (1-\lambda) \eta\left(\Lambda(A+Y^{(2)},L)\right) = b.
        \end{align*}
        % With the same arguments as before it also holds $\eta\left(\Lambda(A+\widetilde{Y},L)\right)\geq b$: otherwise, we could subtract some $\varepsilon > 0$ and make the financial network acceptable with costs strictly less than $\rho_b(G)$. 
        Hence, $\tilde Y$ is an optimal allocation and the set of optimal solutions must be convex.
        \end{enumerate}
        This concludes the proof of Proposition \ref{prop:min}.
\end{proof}

\begin{rem}
  There is indeed no reason to assume that an optimal solution should be unique. Let $\eta(\cdot)=\myE{}{\cdot}$, let $\Lambda$ denote the \textit{total payment shortfall} from Example \ref{exa:lambda:totalpaymentshortfall}, and consider a deterministic financial network $(a,\ell) \in \R_+^3 \times \R_+^{3 \times 3}$ where two nodes 1 and 2 each owe exactly two units of capital to a third node 3. There are no other obligations and all external assets are equal to 1. Without any bailout capital the expected loss of this network is 2. Assume we want to find bailout strategies that reduce the loss to 1. It can easily be seen that all bailout strategies that serve this purpose are $e_1 = (1,0,0)^T$, $ e_2=(0,1,0) ^T$, and all convex combinations of them. For any $\lambda \in (0,1),$
    \begin{align*}
        &\myE{}{\Lambda(a + \lambda e_1 + (1-\lambda) e_2,\ell)}\\ 
        &= \Lambda(a + \lambda e_1 + (1-\lambda) e_2,\ell) = 1- \lambda + 1- (1-\lambda) = 1.
    \end{align*}  
\end{rem}

\section{Computation of systemic risk measures}\label{section:sys_risk_reformulation}
\subsection{Reformulation of systemic risk measures}
In this section, we present a reformulation of the systemic risk measures defined in Section \ref{section:systemic_risk}. This reformulation will motivate for the iterative algorithm for the computation of systemic risk measures that we propose later on.
% In order to formulate an algorithm for the computation of , we introduce an equivalent formulation that will be useful for the numerical computation.
\begin{defn}\label{def:inner_risk}
    Let $G \in \mathcal{G}$ be a stochastic financial network and $c \in \R_+$. We define the inner risk $\rho^I_c(G)$ as the infimum aggregated risk that can be obtained by all capital allocations $Y$ that fulfil the capital constraint $\sum_{n=1}^N Y_n = c$. Then,
    \begin{align}\label{def:inner_risk_formula}
        \rho_c^I(G) := \inf_{Y \in \mathcal{C}^c}\left\{\eta\left(\Lambda(A+Y,L)\right)  \right\},
    \end{align}
    where
    \begin{align*}
        \mathcal{C}^c := \left\{ Y \in L^0(\R_+^N) \,\left|\, \sum_{n=1}^N Y_n = c \right. \right \}.
    \end{align*}
    \end{defn}

\begin{defn}
    With the notion of inner risk from Definition \ref{def:inner_risk}, we call a bailout capital allocation $Y^c \in \mathcal{C}^c$ optimal for capital level $c \in \R_+$ if and only if
    \begin{align*}
        \eta(\Lambda(A+ Y^c,L)) = \rho_c^I(G).    
    \end{align*}
\end{defn} 

\begin{defn}\label{def:outer_risk}
    Let $b >0$. We define the outer risk as
    \begin{align}\label{def:outer_risk_formula}
        \rho_b^O(G) = \inf \left\{c \in \R_+ | \rho_c^I(G) \leq b  \right\}.
    \end{align}
\end{defn}

\begin{lem}\label{lemma2}
    The infimum in (\ref{def:inner_risk_formula}) is attained, hence it is indeed a minimum.
\end{lem}
\begin{proof}
    Let $(Y^{(k)})_{k \in \N} \subseteq \mathcal{C}^c$ be a sequence of allocations such that 
    \begin{align*}
        \eta\left(\Lambda(A+Y^{(k)},L)\right) \xrightarrow[]{k \rightarrow \infty} \rho_c^I(G).
    \end{align*}
    Since the $Y^{(k)}$ are non-negative and sum to $c$, they are also bounded by $c$. Hence, we can use Koml{\'o}s' Theorem \ref{komlos_thm} in the same manner as in (\ref{line:use_komlos}) to obtain a sequence $(\tilde Y^{(k)})_{k \in \N}$, where 
    \begin{align*}
        \tilde Y^{(k)} \in \operatorname{conv}(\{Y^{(k)}, Y^{(k+1)}, \ldots\}),
    \end{align*}
    which lies in $\mathcal{C}^c$ and converges almost surely to a limit $\tilde Y^\infty \in \mathcal{C}^c$.
    Then, also $\Lambda(A+\widetilde{Y}^{(k)},L)$ converges almost surely to $\Lambda(A+\widetilde{Y}^{\infty},L)$.  By the dominated convergence theorem, it also converges in $L^1$, and we can use the continuity of $\eta$ to obtain 
    \begin{align*}
        &\eta\left(\Lambda(A+\widetilde{Y}^{\infty},L)\right) = \eta\left(\Lambda(A+\lim_{k \to \infty} \widetilde{Y}^{(k)},L)\right)\\ 
    &= \eta\left(\lim_{k \to \infty}\Lambda(A+\widetilde{Y}^{(k)},L)\right) = \lim_{k \to \infty}\eta\left(\Lambda(A+\widetilde{Y}^{(k)},L)\right) = \rho_c^I(G).
    \end{align*}
    
    Hence the infimum is attained by $\tilde Y^\infty.$
\end{proof}

\begin{prop}\label{prop:inner_and_outer_risk}
    Let $G \in \mathcal{G}$ and $b > 0$. For the outer risk $\rho_b^O(G)$ in (\ref{def:outer_risk_formula}) and the systemic risk $\rho_b(G)$ in (\ref{def:sys_risk_b}) it holds that
    \begin{align*}
         \rho_b(G) = \rho_b^O(G).
    \end{align*}
\end{prop}

\begin{proof}
    Let $(c^{(k)})_{k \in \N} \subseteq \R_+$ be a sequence of real numbers such that $\rho_{c^{(k)}}^I(G) \leq b$ and $c^{(k)} \xrightarrow[]{} \rho_b^O(G)$.
    
    By Lemma \ref{lemma2}, for every $c^{(k)}$, we obtain an allocation $Y^{(k)} \in \mathcal{C}^{c^{(k)}}$ such that
    \begin{align*}
        \eta\left(\Lambda(A+Y^{(k)},L)\right) \leq b.
    \end{align*}
    Trivially, this sequence fulfils the condition that, for each $k \in \N$, $Y^{(k)} \in \mathcal{C}$ and hence
    these allocations are admissible for bounding the systemic risk $\rho_b(G)$ from above.
    We conclude that $\rho_b(G) \leq \rho_b^O(G).$

    To show the reverse inequality, consider any $Y^* \in \mathcal{C}$ that is optimal for $\rho_b(G).$ This means
    \begin{align*}
        |Y^*|_1 = \rho_b(G),
    \end{align*}
    and by Proposition \ref{prop:min}
    \begin{align*}
       \eta\left(\Lambda(A+Y^*,L)\right) = b.
    \end{align*}
    Then, it follows that
    \begin{align*}
        \rho_b(G) \in \left\{c \in \R_+| \rho_c^I(G) \leq b \right\}. 
    \end{align*}
    Subsequently, it must always be true that
    \begin{align*}
        \rho_b^O(G) \leq \rho_b(G),
    \end{align*}
    which concludes the proof.
\end{proof}
\begin{cor}\label{cor:inner_and_outer_risk}
    The infimum in (\ref{def:outer_risk_formula}) is attained.
\end{cor}
\begin{proof}
    Since $\rho_b(G)$ is attained (Proposition \ref{prop:min}) and $\rho_b(G) = \rho_b^O(G)$ (Proposition \ref{prop:inner_and_outer_risk}) it follows 
    \begin{align*}
        \rho_b^O(G) \in \{c \in  \R_+ | \rho_c^I(G) \leq b\}.
    \end{align*}
\end{proof}
Proposition \ref{prop:inner_and_outer_risk} and Corollary \ref{cor:inner_and_outer_risk} show that, in order to calculate the systemic risk $\rho_b(G)$, we can first find all capital levels $c \in \R_+$ for which the inner risk $\rho_c^I(G)$ is less than or equal to $b$ and then choose the minimum of those values. This procedure will be utilised by the algorithm when approximating the systemic risk of a network.

Next, we turn to the task of finding ``good'' approximations of an optimal allocation $Y^{c} \in \mathcal{C}^c$, for each capital level $c >0$ (for $c = 0$ there is nothing to allocate). As a first step, we show that, under the mild assumption that $\Lambda$ is a measurable function, there exists a measurable function $H^c: \R_{++}^N \times \R_+^{N \times N} \to \R_+^N$ such that $H^c(A,L)$ is optimal for capital level $c >0$. This result is useful, since later we will prove that we can approximate measurable functions arbitrarily close by neural networks.

\begin{prop}\label{prop:function_H_c}
Let  $c >0$, and assume that $\Lambda$ is a measurable function. Then, there exists a measurable function $H^c: \R_{++}^N \times \R_+^{N \times N} \to \R_+^N$ such that 
\begin{align*}
    H^c(A, L) \in \mathcal{C}^c
\end{align*}
is an optimal bailout capital allocation at level $c > 0$.
\end{prop}
\begin{proof}
    Define $E := \{y \in \R_+^N \mid |y| = c\}$ and 
    $h(y,(a, \ell)) := \Lambda(a + y^+, \ell) + \delta_E(y)$,
    where $y^+$ denotes the positive part of $y \in \R^N$ and 
    $$
    \delta_E(y) := \begin{cases}
        0, & \text { if } y \in E\\ 
        +\infty, & \text { otherwise.}
        \end{cases}
    $$
    Then, for fixed $(a,\ell) \in \R^N_{++} \times \R^{N \times N}$, because of the penalty term if $y \notin E$, clearly 
    $$
    \argmin\{h(y,(a, \ell)) \mid y \in \R^N\} = \argmin \{ \Lambda(a+y,\ell) \mid y \in E\}.
    $$
    We show that $h$ is a normal integrand as in Chapter 1.1.2 in \cite{teemu2024convex}, by showing that both $(y, (a,\ell)) \mapsto h_0(y,(a,\ell)) := \Lambda(a+y^+,\ell)$ and $(y, (a,\ell)) \mapsto \delta_E(y)$ are normal integrands. That $\delta_E(\cdot)$ is a normal integrand follows immediately from the example on page 11 in \cite{teemu2024convex}. Regarding $h_0$, we know that for fixed $(a,\ell)$ the function is continuous in $y$. This follows from the fact that $\Lambda$ is convex with respect to its first component on $\R_{++}$ according to Assumption \ref{assume:aggregate} \ref{assume:aggregate:convex/incre}. Since $\Lambda$ is also measurable by assumption, it is in particular measurable in $(a,\ell)$ for fixed $y$. Therefore, $h_0$ is a Carathéodory integrand which is a normal integrand (Example 1.12 in \cite{teemu2024convex}). Since a sum of normal integrands is a normal integrand (\cite{teemu2024convex}, Theorem 1.20) we know by Corollary 1.30 in \cite{teemu2024convex} that $(a,\ell) \mapsto \argmin h(\cdot, (a,\ell))$ admits a measurable selection $H^c: \R_{++}^N \times \R_+^{N \times N} \to E \subset \R_+^N$.
    
    With this measurable selection, it is clear that $H^c(A,L) \in \mathcal{C}^c$ and that $\rho_c^I(G) \leq \eta(\Lambda(A + H^c(A,L),L)).$ Additionally, since $H^c$ minimises $\Lambda(A + H^c(A,L), L))$ pointwise, by the monotonicity of $\Lambda$ and $\eta$, it holds that
    \begin{align*}
        \eta(\Lambda(A + H^c(A,L), L)) \leq \rho_c^I(G).
    \end{align*}
    Therefore, $H^c(A,L)$ is optimal for capital level $c \in \R_+$, which finishes the proof.
\end{proof}
\begin{rem}
    We want to highlight that measurability is a weak condition on $\Lambda$. For example, if $\Lambda$ is one of the aggregation functions from Examples \ref{exa:lambda:totalpaymentshortfall}, \ref{exa:lambda:2}, or \ref{exa:lambda:3}, we know that $\Lambda$ is measurable as a composition of measurable functions if the clearing vector function $\overline{\operatorname{CV}}$ from Definition \ref{def:CV} is measurable. However, from \cite{EisNoe} we know that the clearing vector $\overline{\operatorname{CV}}$ can be obtained by applying $\Phi$ from (\ref{line:phi}) iteratively:
    $$
    \overline{\operatorname{CV}}(a, \pi,\bar \ell) = \lim_{n \to \infty}\Phi(\cdot, a, \pi, \bar \ell)\, \underbrace{\circ \cdots \circ}_{n \text{ times}}\, \Phi(0, a, \pi, \ell).
    $$
    Therefore, $\overline{\operatorname{CV}}$ is measurable as a limit of measurable functions.
\end{rem}

\subsection{Iterative optimisation algorithm}\label{chapter:comp_theory}

Computing the systemic risk is not straightforward due to the scenario-dependent allocation. However, the results from Proposition \ref{prop:inner_and_outer_risk} allow us to formulate an iterative optimisation algorithm based on the empirical risk.
% \begin{Notes}{Note}
\begin{defn}
    Given a risk measure $\eta : L^1(\Omega, \mathcal{F}, \mathbb{P}; \R) \to \R$ and a random variable $X \in L^1(\R)$ we define the empirical risk $\hat \eta$ of $n\in \N$ i.i.d. realisations $X^1,\ldots,X^n \in \R$ of $X$ as 
    \begin{align*}
        \hat \eta(X^1,\ldots,X^n) := \eta(\tilde X)
    \end{align*}
    where $\tilde X$ is a random variable that is uniformly distributed on the realisations, i.e., $\mathbb{P}(\tilde X = X^i) = \frac{1}{n}$ for all $i \in [n]$.
\end{defn} 
The general idea of the iterative optimisation algorithm is the following. For every iteration, we randomly generate samples of the stochastic financial network. An inner optimisation tries to minimise the empirical aggregated risk $\hat \eta$ by allocating the current bailout capital $c_{curr} \in \R$ optimally. An outer optimisation checks whether the empirical aggregated risk obtained from the bailout capital allocation is higher or lower than the acceptable risk threshold $b$, increases or decreases $c_{curr} \in \R$ accordingly, and hands either back to the inner optimisation or terminates the algorithm when the maximum number of iterations is reached.\\

More in detail, from Proposition \ref{prop:function_H_c} we know that, for any capital level $c \in \R_+$, there exists a measurable function $H^c: \R_{++}^N \times \R_+^{N \times N} \to \R_+^{N}$ such that $H^c(A,L)$ is an optimal bailout capital allocation, in the sense that it minimises the inner risk. Since $c=0$ indicates that there is no capital to be allocated, we focus on the case $c>0$ and consider the normalised function
\begin{align*}
    \varphi^c = \frac{H^c}{c}
\end{align*}
such that $H^c(A,L) = c  \varphi^c(A,L).$
This separation into ``amount of bailout capital'' and ``distribution of capital'' is motivated by the algorithm. Overall, we aim to approximate $H^c(A,L)$. But instead of approximating $H^c(A,L)$ directly for each level of bailout capital, we approximate $\varphi^c(A,L)$ by a parameterised function $\phi^\theta,$ where $\theta \in \R^m$ is a collection of parameters for some $m\in \N$. Since $\varphi^c(A,L)$ sums to one for all $c \in \R$, in the implementation we can use a softmax layer. Furthermore, we can use the optimal parameter set corresponding to bailout capital level $c_{\text{old}}$ as the initial guess for the parameter set of the new bailout capital level $c_{\text{new}}$ in the next iteration.

To find the optimal parameter set ${\theta^c}^*$, we use i.i.d. samples from the distribution of the stochastic financial network $(A,L)$ and minimise the empirical aggregated risk
\begin{align}\label{empRisk}
% J_I(\theta, [A^i,L^i]_{i\in[n]}, c)  := 
\hat\eta\left(\left[\Lambda(A^i + c \phi^\theta(A^i,L^i), L^i)\right]_{i \in [n]}
% ,\hdots ,\Lambda(A^n + c\phi^\theta(A^n,L^n), L^n)
\right).
\end{align}
In particular, in the case $\eta = \myE{}{\cdot}$, the empirical aggregated risk is given by
\begin{align*}
   \frac{1}{n}\sum_{i=1}^n\Lambda(A^i + c \phi^\theta(A^i,L^i),L^i).
\end{align*}
By the proof of Proposition \ref{prop:function_H_c}, we see that $H^c$ minimises the inner risk by minimising $\Lambda$ pointwise. This means that the inner optimisation can be carried out pointwise, too. Therefore, different choices for the loss function, other than the empirical risk, are possible for this step. In order to find the optimal parameters, we minimise this objective function in mini-batches with stochastic gradient descent, specifically the Adam optimiser from \cite{kingma2014adam}, which is well-established in the machine learning community.\\

In the outer optimisation, we want to find the smallest capital level where the inner risk equals $b$ (see Proposition \ref{prop:min}), or equivalently, we are searching for a root of the function $c \mapsto \eta(\Lambda(A + c\phi^{{\theta^c}^*}(A,L), L)) -b$. Since $\eta$ is monotone deceasing with respect to the bailout capital level, we could use a simple bisection search if it were possible to exactly evaluate the inner risk for the current capital level $c$. Unfortunately, we cannot calculate the exact inner risk because the optimal parameter set ${\theta^c}^*$ is unknown, and we are not necessarily working on a finite probability space. We can only rely on the parameter set $\theta$ obtained from the inner optimisation and calculate the empirical aggregated risk of the samples that we randomly generated for the current iteration from (\ref{empRisk}).
Therefore, a deterministic bisection search is risky. In theory, one adverse set of samples could result in excluding the interval containing the true root. In practice, an even bigger problem is that, most likely, after one iteration the bailout capital allocation is not optimal yet. A suboptimal bailout capital allocation increases the aggregated risk and can result in discarding the interval with the true root. Therefore, we rely on the \textit{probabilistic bisection search}, which can handle the situation when the evaluation of the function is distorted by random errors \cite{frazier2019probabilistic}, \cite{horstein1963sequential}. 
This version of the bisection search can approximate a root $c^* \in [c_\text{min},c_\text{max}]$ of a monotone decreasing objective function 
% $g:\R \to \R$
based on noisy observations. The basic idea is to start with a uniform density $f_0$ on $[c_\text{min},c_\text{max}]$ and then iteratively update the density $f_j$ from noisy observations of the objective function at the median of $f_{j-1}$. The resulting algorithm, where we combine inner and outer optimisation in order to approximate the systemic risk, is summarised in Algorithm \ref{Algo:1}.\\
% \end{Notes}

\begin{algorithm}
\caption{Algorithm to approximate minimal bailout capital and its optimal allocation}
\label{Algo:1}
\begin{algorithmic}[1]
\Require 
\State Minimal bailout capital: $c_\text{min}$
\State Maximal bailout capital: $c_\text{max}$
\State Number of epochs: $N_\text{epochs}$
\State Number of samples: $N_\text{samples}$
\State Risk threshold: $b$
\State Bisection parameters $p\in [\frac{1}{2},1)$, $q=1-p$
\State \textbf{Initialisation:}
\State \quad Initialise $f_0$ as uniform distribution on $[c_\text{min}, c_\text{max}]$
\State \quad Initialise parameterised function: $\phi^{\theta_0}$
\State \textbf{Iteration:}
\For{$j=1,\ldots,N_\text{epochs}$}
    \State Compute 
    $
    c_j = F_j^{-1}(1/2)
    $
    \State Sample i.i.d. realisations $(A^1,L^1),\ldots,(A^{N_\text{samples}},L^{N_\text{samples}})$ of $(A,L)$
    \State Compute 
    $
    J(\theta_j)
    = \hat\eta\left([\Lambda(A^i + c_j \phi^{\theta_j}(A^i,L^i), L^i)]_{i \in [N_\text{samples}]}\right).
    $
    \State Compute $\nabla_{\theta_j} J$
    \State Compute ${\theta_{j+1}}$ with Adam optimiser
    \State Compute $f_{j+1}:$
    \begin{align*}
        \text{if $J > b$, then } f_{j+1}(x)= \begin{cases}
        2 p f_j(x), & \text { if } x \geq c_j, \\ 
        2 q f_j(x), & \text { if } x<c_j,
        \end{cases}\\
        \text{if $J \leq b$, then }f_{j+1}(x)= 
        \begin{cases}2 q f_j(x), & \text { if } x \geq c_j, \\
        2 p f_j(x), & \text { if } x<c_j .\end{cases}
    \end{align*}
    \EndFor
    \State \textbf{Output:} $c_{N_\text{epochs}},\phi^{\theta_{N_\text{epochs}}}$ 
\end{algorithmic}
\end{algorithm}

If the approximation of the optimal bailout allocations works sufficiently well for every capital level $c \in \R_+$, the approximated systemic risk obtained in the optimisation will be close to the actual systemic risk $\rho_b(G)$. Furthermore, even if the optimal allocations are not solved accurately, the lowest capital level $c \in \R_+$ that we find is, in any case, an estimated upper bound for $\rho_b(G)$. Additionally, $c \phi^\theta$ offers a bailout strategy that secures the system. Hence, even rather `na\"ive'' allocations $c\phi^\theta$ can yield some valuable information. They provide a value $c \in \R_+$ that suffices to secure the system and provide instructions on how to allocate the capital in each scenario.

\begin{rem}
    It is worth pointing out that the traditional approach (\ref{line:sys_risk_vec_det_alloc}) of a deterministic allocation of bailout capital $(m_1,\ldots,m_N) \in \R^N$, i.e., an allocation that does not depend on scenario $\omega \in \Omega$, is also covered by this setting. In that situation, $\phi^{\theta}$ would not parametrise a function that maps $(a, \ell) \in \R_{++}^N \times \R_+^{N \times N}$ to an optimal distribution, but would instead directly represents the scenario-agnostic distribution
\begin{align*}
    \phi^{\theta}(a,\ell) = \phi^{\theta} = (\theta_1,\ldots,\theta_N)^T
\end{align*}
that maps every scenario $(a, \ell)$ to the same constant bailout distribution that minimises the systemic risk of $G$.
\end{rem}

\begin{rem}
    If the probability space is finite and small and the loss function sufficiently simple, one might not need the bisection procedure and can instead formulate a single convex optimization problem that simultaneously determines the minimal bailout capital and its optimal allocation across all scenarios. For instance, when using the expectation risk measure and the Eisenberg–Noe based \textit{total payment shortfall} from Example \ref{exa:lambda:totalpaymentshortfall} as loss function, the problem reduces to a linear program due to the linear programming characterization of the Eisenberg-Noe clearing mechanism and can be solved directly by linear programming. However, even in this linear setting, memory requirements grow rapidly with the size of the probability space rendering these approaches impractical for larger spaces. For more challenging convex problems, large finite probability spaces, and general probability spaces, traditional optimization techniques (for example the approach in  \cite{ararat2023computation}) become computationally prohibitive. In this situation, our proposed methodology offers a scalable and feasible alternative that allows to solve large convex optimisation problems with an improved accuracy due to the high number of data points. Since we do not require access to all data points simultaneously, our method is suitable for large datasets of arbitrary size and, in particular, general probability spaces from which we draw new samples in each epoch without impractical memory requirements.
\end{rem}

In the next section, we discuss several types of neural networks that we use as $\phi^\theta$ in our algorithm for the computation of systemic risk measures. 

%###################################################

\section{Approximation via neural networks}\label{section:neural_networks}
In this section, we introduce three different neural network architectures that we use to approximate the function that maps a financial network to an optimal bailout allocation, as in Proposition \ref{prop:function_H_c}.
More precisely, in addition to classical feedforward neural networks (FNNs) and graph neural networks (GNNs), we present the class of permutation-equivariant neural networks (PENNs), which were first described in \cite{herzig2018mapping}. We extend this architecture by a modification that we call extended permutation-euqivariant neural network (XPENN). These XPENNs unite favourable properties of GNNs and PENNs, and perform well in the numerical experiments of Section \ref{section:numerical_experiments}, where we compare all these neural network architectures.

\subsection{Feedforward neural networks}
Feedforward neural networks (FNNs) are the original and simplest type of neural networks. They have been successfully applied to a wide range of problems and are typically defined as follows.
\begin{defn}\label{def:NN}
    Let $L \in \N$ be the number of layers and $d_0,\ldots,d_L \in \N$ corresponding layer dimensions. Let further $\sigma: \R \to \R$ be a Borel measurable function and $W_l:\R^{d_{l-1}}\to\R^{d_l}$ be an affine function for all $l \in [L].$ Then, a  \emph{feedforward neural network} $F:\R^{d_0} \to \R^{d_L}$ is defined by
    \begin{align*}
        F = W_L \circ F_{L-1}\circ  \dotsb \circ F_1
    \end{align*}
    where $F_l = \sigma \circ W_l$ for $l=1,\ldots,L-1$ and the activation function $\sigma$ is applied componentwise.
\end{defn}
Motivated by their success in applications, their theoretical properties have also been extensively studied. An important starting point is the class of universal approximation theorems for learning functions, see \cite{cybenko1989approximation, hornik1991approximation, leshno1993multilayer}. For example, it is a well-known result that neural networks are universal approximators for measurable functions -- see, for example, the following result from \cite{leshno1993multilayer}, as formulated in \cite{biagini2023neural}.

\begin{prop}[Theorem B.1, \cite{biagini2023neural}] \label{thm:universality_of_measurable_thomas}
    Assume the activation function $\sigma$ to be bounded and non-constant. Let $f:(\R^d, \mathcal{B}(\R^d)) \to (\R^m, \mathcal{B}(\R^m))$ be a measurable function and let $\mu$ be a probability measure on $(\R^d, \mathcal{B}(\R^d))$. Then, for any $\varepsilon, \bar \varepsilon >0$, there exists a neural network $g$ such that
    \begin{align*}
        \mu ( \{x \in \R^d: |f(x) - g(x)|_p \geq \bar \varepsilon\}) \leq \varepsilon.
    \end{align*}
\end{prop}
\begin{proof}
    See \cite{biagini2023neural}.
\end{proof}
This shows that, theoretically, feedforward neural networks can approximate any measurable function in probability.
However, as we will see later, they do not perform very well in our experiment on the approximation of functions based on graph data -- such as $H^c$ from Proposition \ref{prop:function_H_c}.

\subsection{Graph neural networks}
Graph neural networks (GNNs) are a family of neural networks that are specifically designed to operate on graph-structured data. We are interested, in particular, in spatial graph neural networks, where information is gathered locally by message passing, i.e., the process of passing information between the nodes of a graph along its edges. Their origins lie in \cite{scarselli_new}, \cite{scarselli_gori_comp_capabilities_of_GNNs}, while more current literature includes \cite{gilmer2017neural} and the references therein, as well as
 \cite{hamilton2017inductive}, \cite{kipf2016semi}, \cite{NEURIPS2022_GNN_MPNN_Generalization_Gitta}, \cite{velivckovic2017graph}, and \cite{jegelka2018powerful}. The basic idea in these GNNs is that each node in the graph is associated with a vector representation (often also called feature or hidden-state vector), and at each iteration of message passing, the vector representation of a node is updated by aggregating the representations of its neighbouring nodes and then combining them with the node's previous representation to derive an updated vector representation. 
\begin{defn}\label{def:neighbourhood_agg}
    A function 
    \begin{align*}
        f: \bigcup_{k \in \N} \left(\R^{m_h} \times \R^{m_e} \right)^k \to \R^{m_{h}'},
    \end{align*} that maps any number $k \in \N$ of node and edge features $(h_1,e_1),\ldots,(h_k,e_k) \in \R^{m_h} \times \R^{m_e}$ of dimension $m_h \in \N$ and $m_e \in \N$, respectively, to some aggregated representation vector in $\R^{m_h'}$ of dimension $ m_h' \in \N$, and is permutation invariant in the sense of Definition \ref{def:permutation_invari}, is called an aggregation function.\footnote{Definition \ref{def:permutation_invari} is applicable if we identify $\R^{m_h} \times \R^{m_e}$ with $\R^{m_h + m_e}$.}
\end{defn}
In GNNs, such aggregation functions are used to aggregate the features of neighbouring nodes and the edges between them into an aggregated feature (or representation vector) of the neighbourhood.

Recalling that $\mathcal{D}$ denotes the domain of graphs, $\mathcal{D} = \R^{N \times d} \times \R^{N \times N \times d'}$ from Definition \ref{def:domain_of_graphs}, a GNN is defined as follows.
\begin{defn}\label{def:GNN}
    Let $L \in \N$ be the number of iterations, $m_1,\ldots,m_L \in \N$ the dimensions of the corresponding node representations, and $m'_1,\ldots,m'_L \in \N$ the dimensions of the neighbourhood representations in the respective iteration. Recall that $d,d' \in \N$ are the dimensions of the initial node and edge representations and $N \in \N$ the graph size of graphs in $\mathcal{D}$.
    Choose $m_0=d$ and let 
    \begin{align*}
    f^{(l)}_{ag}:\bigcup_{k \in \N}\left(\R^{m_{l-1}} \times \R^{d'} \right)^k 
    % \{\{(h_1,e_1),\ldots,(h_k,e_k)\}| k \in \N, (h_1,e_1),\ldots,(h_k,e_k) \in \R^{m_{l-1}}\times \R^{d'}\} 
    \to \R^{m'_{l}},
    \quad l= 1,\ldots,L,
    \end{align*} 
    be aggregation functions as in Definition \ref{def:neighbourhood_agg}.
    % that, in every iteration $l= 1,\ldots,L$, map any set of node and edge representations to some aggregated neighbourhood representation.
    Furthermore, let 
    $$f^{(l)}_{com}: \R^{m_{l-1}} \times \R^{m'_l} \to \R^{m_l},
    \quad l= 1,\ldots,L,
    $$ be functions that map a node representation and a neighbourhood representation to an updated node representation.\\
    For every node $i\in [N]$, let $h_i^{(l)}$ be its node representation in iteration $l \in [L].$ The node features $a_i$ serve as the initial node representations  
    $$
    h_i^{(0)} = a_i,
    $$
    and the edge representations, which are not updated in this type of GNN, are given by the edge features, i.e., for any nodes $i,j\in [N]$ 
    $$
    e_{ij} = \ell_{ij}.
    $$
    Then, at every iteration step $l=1,\ldots,L-1$, the new node representation $h_i^{(l)}$ is calculated as
    $$
    h_i^{(l)} = f_{act}(f^{(l)}_{com}(h_i^{(l-1)}, f^{(l)}_{ag}(\{(h_j^{(l-1)}, e_{ji}) \mid j \in N(i)\}))),
    $$
    where $f_{act}:\R \to \R$ is some non-linear activation function applied componentwise. In the last step, the activation is not applied, such that
    \begin{align*}
        h_i^{(L)} = f^{(L)}_{com}(h_i^{(L-1)}, f^{(L)}_{ag}(\{(h_j^{(L-1)}, e_{ji}) \mid j \in N(i)\})).
    \end{align*}
    A function $f: \mathcal{D} \to \R^{N \times m_L}$ defined as
    \begin{align*}
        f(a,\ell) = \begin{pmatrix}
            h^{(L)}_1\\\vdots\\h^{(L)}_N
        \end{pmatrix}
    \end{align*}
    is called a \emph{graph neural network} with $L$ rounds of message passing.
\end{defn}

\begin{rem}
    In comparison to feedforward neural networks (Definition \ref{def:NN}), it is not obvious from the above definition why such a function is called a graph \textit{neural network}. The reason for this name is that the functions $f^{(l)}_{ag}$ and $f^{(l)}_{com}$ are usually chosen as feedforward neural networks with trainable parameters. One example is the architecture we choose for our numerical experiments, which is presented in Definition \ref{def:SAGEConv}. 
\end{rem}

The specific GNN architecture that we consider is called SAGEConv (with mean aggregation) and was introduced in \cite{hamilton2017inductive}.
\begin{defn}\label{def:SAGEConv}
    With the notation from Definition \ref{def:GNN}, we call a GNN layer a \textit{SAGEConv} layer if the aggregation function is given by
    \begin{align*}
        f_{ag}(\{(h_j^{}, e_{ji}) \mid j \in N(i)\}) := \frac{1}{|N(i)|} \sum_{j \in N(i)} e_{ji}h_j,
    \end{align*}
    and denoting 
    \begin{align*}
        h_{N(i)} =  f_{ag}(\{(h_j^{}, e_{ji}) \mid j \in N(i)\}),
    \end{align*}
    the combination function is
    \begin{align*}
        f_{comb}(h_i, h_{N(i)}) := W \cdot \operatorname{concat}(h_i, h_{N(i)}),
    \end{align*}
    where $W$ is the representing matrix of some linear function of appropriate dimension, and $\operatorname{concat}(h,h') \in \R^{n+n'}$ is the concatenation of two vectors $h \in \R^n$ and $h' \in \R^{n'}$ of any dimensions $n,n' \in \N$.
    As activation function, we utilise the sigmoid activation
    \begin{align*}
        f_{act}(x) := \frac{1}{1+e^{-x}}.
    \end{align*}
\end{defn}
There are many other possible choices regarding GNN layers. One important reason for using the GNN in Definition \ref{def:SAGEConv} is that its message passing also accounts for edge weights. 
Other popular GNN layers that have this property include the GIN layer from \cite{jegelka2018powerful} and the NNConv layer from \cite{gilmer2017neural}. However, comparing multiple GNN architectures would exceed the scope of this work.

The most important reason why GNNs are a suitable design for approximating the target function $H^c$ from (\ref{prop:function_H_c}) is their permutation equivariance. Specifically, if $y \in \R^N$ is an optimal bailout capital allocation for a financial network $(a, \ell) \in \mathcal{D},$ then $\sigma(y)$ is optimal for $\sigma(a,\ell),$ for any permutation $\sigma \in S(N).$ Hence $H^c$ is permutation-equivariant, see Definition \ref{def:permutation_equi}. Since GNNs calculate the message that reaches a node $i$ based on its neighbourhood $N(i)$, the ordering of the nodes does not matter as long as the network topology remains unchanged. This means that the output calculated for each node during multiple steps of message passing is permutation-equivariant by design:
For $\sigma \in S_N$, the GNN $f: \mathcal{D} \to \R^{N \times l}$ automatically computes results that respect
$$
f(\sigma(g)) = \sigma(f(g)).
$$
The situation is very different for feedforward neural networks. For example, let $f:\R^{N \times m} \to \R^{m'}$ be a neural network \footnote{We identify $\R^{N \times m}$ with $\R^{N \cdot m}$ to be in the setting of Definition \ref{def:NN}.} and let $\sigma \in S_N$ be any permutation. Then, in general, for some $x \in \R^{N \times m}$, the equality 
$$
f(\sigma(x)) = \sigma(f(x))
$$
does not hold. Since a feedforward neural network is not permutation-equivariant by design, it needs to learn this property during training when approximating $H^c$, which makes the learning task more complicated.

\subsection{Permutation-equivariant neural networks}
Motivated by the conjecture that permutation equivariance could be a crucial advantage of GNNs over FNNs for the approximation of the target function $H^c$, we present a neural network architecture that lies at the border between FNNs and GNNs. We will call this neural network architecture a \textit{permutation-equivariant neural network} (PENN). 
Such PENNs do not utilise iterated message passing. However, they process the information given by a graph in a permutation-equivariant manner and  ``provide a summary of the entire graph,'' as stated in \cite{herzig2018mapping}, where this architecture was first studied. Furthermore, we are able to recover a universal approximation result, which is not available for GNNs.

Before we introduce PENNs, we investigate permutation-equivariant functions in more detail. As outlined in the example regarding equations (\ref{line:repr_not_unique_1}) and (\ref{line:repr_not_unique_2}), for a graph there exist multiple possible representations $g_1,g_2,\ldots \in \mathcal{D}$ that are equally valid. Each of these representations implies the choice of some node order, in the sense that we chose $a_1$ to be the feature of the ``first node'' and $a_N$ to be the feature of the ``$N$-th node''. We can incorporate this (or any other) choice of numbering into the node features by considering extended node features $\widetilde a$, where
\begin{align*}
    \widetilde a = \begin{pmatrix}
        a_1, 1\\ \vdots \\ a_N, N
    \end{pmatrix}.
\end{align*}

Adding a node ID as a feature does not appear very restrictive since it only highlights an arbitrary choice that was already made earlier by ordering the nodes. This motivates the following definition of the domain of ID-augmented graphs.

\begin{defn}
    Let $\bar \sigma = (\sigma(1),\ldots,\sigma(N))^T$ be the vector obtained by evaluating some permutation $\sigma \in S(N)$ on the numbers $1,\ldots,N$. 
    The ID-augmented graph domain $\widetilde{\mathcal{D}}$ contains any element of $\mathcal{D}$ in which the node features $a$ are augmented with an ordering $\bar \sigma$:
    \begin{align*}
        \widetilde{\mathcal{D}} := \left\{ (\widetilde a, \ell)\,\left|\, \widetilde a = (a, \bar \sigma), \sigma \in S(N), (a,\ell) \in \mathcal{D}\right.\right\}.
    \end{align*}
\end{defn}

\begin{rem}
    Since the augmented node ID could simply be interpreted as one additional node feature, the ID-augmented graphs $\widetilde{\mathcal{D}}$ with node feature size (before augmentation) $d \in \N$ are a subset of graphs with node feature size $d+1$. Therefore, the definition of permutation-equivariant functions (Definition \ref{def:permutation_equi}) extends naturally to $\widetilde{\mathcal{D}}.$ 
\end{rem}

The following result is a modified version of Theorem 1 in \cite{herzig2018mapping}. We rigorously prove that any permutation equivariant function on the domain of directed, weighted, and featured graphs, whose node features are augmented with node IDs, exhibits a certain structure. Based on this structure, we later define a permutation-equivariant neural network architecture using feedforward neural networks.
\begin{prop}\label{prop:repr_of_GPEqui}
         A function $\widetilde{\tau}:\widetilde{\mathcal{D}} \to \R^N$ is permutation-equivariant as in Definition \ref{def:permutation_equi} if and only if there exist dimensions $V \leq N^2d'$, $W \leq N(d+1)+ N^2d'$, and functions $\varphi: \mathbb{R}^{2 (d+1)+d'} \rightarrow \mathbb{R}^{V}$, $\alpha: \mathbb{R}^{d+1+V} \rightarrow \mathbb{R}^W$, and $\rho: \mathbb{R}^{d+1+W} \rightarrow \mathbb{R}$ such that for all $(\widetilde{a},\ell) \in \widetilde{\mathcal{D}}$ and for all $k=1, \ldots, N$:
     \begin{align*}
         \widetilde{\tau}(\widetilde{a},\ell)_k =\rho\left(\tilde a_k, \sum_{i=1}^N \alpha\left(\tilde a_i, \sum_{j \neq i} \varphi\left(\tilde a_i, \ell_{i, j}, \tilde a_j\right)\right)\right).
     \end{align*}
\end{prop}

\begin{proof}
    First, we show that any $\widetilde{\tau}$ satisfying the R.H.S of (\ref{line:only_PI_on_tilde_D}) is permutation-equivariant.
    Let $V$, $W$, $\varphi$, $\alpha$, and $\rho$ be given and let $\gamma \in S(N)$ be a permutation of $[N]$. Then, we need to show that
    \begin{align*}
        \widetilde{\tau}\left(\gamma \left(\widetilde{a},\ell\right)\right)_k = \gamma \left(\widetilde{\tau}\left(\widetilde{a},\ell\right)\right)_k.
    \end{align*}
    It holds by definition that
    \begin{align*}
        \widetilde{\tau}\left(\gamma \left(\widetilde{a},\ell\right)\right)_k &=  \widetilde{\tau}\left(\gamma (\widetilde{a}), \gamma(\ell)\right)_k\\
        &=\rho\left(\gamma(\widetilde{a})_k, \sum_{i=1}^N \alpha\left(\gamma(\widetilde{a})_i, \sum_{j  \neq i} \varphi\left(\gamma(\widetilde{a})_i, \gamma(\ell)_{ij}, \gamma(\widetilde{a})_j\right)\right)\right).
    \end{align*}
    Further, since $\gamma(\widetilde{a})_k = \widetilde{a}_{\gamma^{-1}(k)}$ and $\gamma(\ell)_{ij} = \ell_{\gamma^{-1}(i),\gamma^{-1}(j)}$, we have that
    \begin{align*}
        &\rho\left(\gamma(\widetilde{a})_k, \sum_{i=1}^N \alpha\left(\gamma(\widetilde{a})_i, \sum_{j  \neq i} \varphi\left(\gamma(\widetilde{a})_i, \gamma(\ell)_{ij}, \gamma(\widetilde{a})_j\right)\right)\right)\\
        &=\rho\left(\widetilde{a}_{\gamma^{-1}(k)}, \sum_{i=1}^N \alpha\left(\widetilde{a}_{\gamma^{-1}(i)}, \sum_{j  \neq i} \varphi\left(\widetilde{a}_{\gamma^{-1}(i)}, \ell_{\gamma^{-1}(i),\gamma^{-1}(j)}, \widetilde{a}_{\gamma^{-1}(j)}\right)\right)\right)\\
        &=\rho\left(\widetilde{a}_{\gamma^{-1}(k)}, \sum_{i=1}^N \alpha\left(\widetilde{a}_{i}, \sum_{j  \neq i} \varphi\left(\widetilde{a}_{i}, \ell_{ij}, \widetilde{a}_{j}\right)\right)\right)\\
        &= \widetilde{\tau}(\widetilde{a},\ell)_{\gamma^{-1}(k)},
    \end{align*}
    where we used the fact that $\{1,\ldots,N\} = \{\gamma(1),\ldots,\gamma(N)\}$.
    This shows that a function defined by the right-hand side in (\ref{line:only_PI_on_tilde_D}) is indeed permutation-equivariant on $\widetilde{\mathcal{D}}$, for any $\varphi$, $\alpha$, and $\rho$. 
    Next, we prove that any given permutation-equivariant function $\widetilde{\tau}$ can be expressed by carefully chosen $\varphi$, $\alpha$, and $\rho$ in (\ref{line:only_PI_on_tilde_D}). 
    The key idea is to construct $\varphi$, and $\alpha$, such that the second argument of $\rho$ contains all the information about the graph $(\widetilde{a},\ell)$. Then, the function $\rho$ corresponds to an application of $\widetilde{\tau}$ to this representation. To simplify notation, we assume node and edge features are scalar $(d=d'=1)$. The extension to vectors is simple, but involves more indexing.

    Let $(\widetilde{a},\ell)=((a, \bar \sigma),\ell) \in \widetilde{\mathcal{D}}$. We define $\varphi$ to return a $N \times N$ zero matrix with only one entry that is allowed to be non-zero, which is the value $\ell_{ij}$ at index $(\bar {\sigma}_i, \bar {\sigma}_j) = (\sigma(i),\sigma(j))$. Then, with $\mathds{1}_{\sigma(i),\sigma(j)}^{N \times N}$ denoting an $N \times N$ matrix that is one at index $(\sigma(i),\sigma(j))$ and zero elsewhere,  
        \begin{align*}
                \varphi(\tilde a_i, \ell_{ij}, \tilde a_j) = \varphi((a_i,\sigma(i)), \ell_{ij}, (a_j,\sigma(j))) = \mathds{1}_{\sigma(i),\sigma(j)}^{N \times N}\cdot \ell_{ij}.
    \end{align*} 
    This means that the function $\varphi$ stores the information about the edge from node $i$ to $j$ in position $(\sigma(i),\sigma(j)).$ After summing over all $j \neq i$, the resulting matrix
    \begin{align*}
         M_\varphi^i = \sum_{j \neq i}\varphi(\tilde a_i, \ell_{ij}, \tilde a_j)
    \end{align*}
    contains all information about the outgoing edges of any node $i$.

    Next, we define the function $\alpha$ to augment this matrix with an $N\times2$ matrix that contains
    $a_i, \sigma(i)$ in row $\sigma(i)$, 
    and zeros elsewhere.
    \begin{align*}
        \alpha(\tilde a_i, M^i_\varphi) =  \alpha((a_i,\sigma(i)), M^i_\varphi) = \left( \left(\mathds{1}_{\sigma(i),1}^{N \times 1} \cdot a_i,\mathds{1}_{\sigma(i),1}^{N \times 1} \cdot \sigma(i)\right), 
        % \mathds{1}_{\sigma(i),:}^{N \times 2} \cdot (a_i,\sigma(i)), 
        M_\varphi\right).
    \end{align*}
    The resulting tuple generated by $\alpha$ contains the node feature of node $i$, its node ID $\sigma(i)$, and all information about its outgoing edges. Summing all terms generated by $\alpha$, for $i=1, \ldots, N,$ yields
    \begin{align*}
        \sum_{i=1}^N\alpha(\tilde a_i, M^i_\varphi) = (\sigma(\widetilde{a}) , \sigma(\ell)) = \sigma(\widetilde a, \ell).
    \end{align*}

    In the final step, we apply $\widetilde{\tau}$ to $\sigma(\widetilde{a},\ell)$. For node $k$ we want to return the $k$-th component of $\widetilde{\tau}(\widetilde{a},\ell)$ which corresponds to the $\sigma(k)$-th component of $\sigma(\widetilde{\tau}(\widetilde{a},\ell))$, by permutation invariance. Hence, we choose $\rho$ such that
    \begin{align*}
        \rho\left(\tilde a_k, \sigma(\widetilde{a},\ell))\right) = \rho\left((a_k,\sigma(k)), \sigma(\widetilde{a},\ell)\right) =  \widetilde{\tau}(\sigma(\widetilde{a},\ell))_{\sigma(k)} = \widetilde{\tau}(\widetilde{a},\ell)_k.
    \end{align*}
    This concludes the proof.
\end{proof}

This theoretical result motivates us to introduce a certain architecture of neural networks that was already described in \cite{herzig2018mapping} and is defined as follows.

\begin{defn}\label{def:PENN}
    Let $\hat\rho, \hat\alpha, \hat\varphi$ be feedforward neural networks $\hat\varphi: \R^{2d+d'} \to \R^{d_1}, \hat\alpha: \R^{d+d_1}\to \R^{d_2},  \hat\rho: \R^{d+d_2}\to \R,$ for dimensions $d_1,d_2\in \N$.
    The function $f: \R^{N \times d} \times \R^{N \times \N \times d'} \to \R^N$ defined componentwise for any $k \in [N]$ as 
    \begin{align*}
        f(a,\ell)_k =
        \hat\rho\left(a_k, \sum_{i=1}^N \hat\alpha\left(a_i, \sum_{j \neq i} \hat\varphi\left(a_i, \ell_{ij},a_j\right)\right)\right)
    \end{align*}
    is called \emph{permutation-equivariant neural network} (PENN).
\end{defn}

The next proposition provides a universal approximation result for PENNs introduced in Definition \ref{def:PENN}.
We show that, by approximating $\rho$, $\alpha$, and $\varphi$ with neural networks $\hat \rho$, $\hat \alpha$, and  $\hat \varphi$, we can, in probability, approximate the composition of $\rho$, $\alpha$, and $\varphi$ arbitrarily well by the composition of $\hat \rho$, $\hat \alpha$, and $\hat \varphi$, given a probability measure on $\widetilde{\mathcal{D}}$. The result we show is even more general, because it holds for functions and probability measures on $\R^{N \times (d+1)} \times \R^{N \times N \times d'} \supset \widetilde{\mathcal{D}}$. 

\begin{prop}\label{prop:universality_of_composition}
    Let $\rho$, $\alpha$, and $\varphi$ be Borel measurable functions $\varphi: \R^{2d_1+d_2} \to \R^{d_3}$, $\alpha: \R^{d_1+d_3}\to \R^{d_4}$, and $\rho: \R^{d_1+d_4}\to \R$, for arbitrary dimensions $d_1,\ldots,d_4\in \N$.
    Define $g:\R^{N \times d_1}\times \R^{N\times N \times d_2} \to \R^N$ for all $k \in [N]$ by
    \begin{align*}
       g(x,y)_k :=
           \rho\left(x_k,\sum_{i=1}^N\alpha(x_i,\sum_{j \neq i}^N \varphi(x_i, y_{ij},x_j))\right).
    \end{align*}
    Then, for any probability measure $\mu$ on $(\R^{N \times d_1}\times \R^{N\times N \times d_2}, \mathcal{B}(\R^{N \times d_1}\times \R^{N\times N \times d_2}))$, any $p \in [1,\infty]$, and any $\varepsilon, \bar \varepsilon>0$, there exist feedforward neural networks $\hat \rho$, $\hat \alpha$, and  $\hat \varphi$ such that for the function $\hat g:\R^{N \times d_1}\times \R^{N\times N \times d_2} \to \R^N$ defined by
    \begin{align*}
        \hat g (x,y)_k = 
           \hat\rho\left(x_k,\sum_{i=1}^N\hat\alpha(x_i,\sum_{j \neq i}^N \hat\varphi(x_i, y_{ij},x_j))\right), \quad k \in [N],
    \end{align*}
    it holds
    \begin{align*}
        \mu \left( \left. \big\{(x,y) \in \R^{N \times d_1}\times \R^{N\times N \times d_2} \,\right|\, |g(x,y) - \hat g(x,y)|_p \geq \bar \varepsilon  \big\}\right) \leq \varepsilon.
    \end{align*}
\end{prop}
\begin{proof}
    Let $\varepsilon,\bar \varepsilon>0$.
    Choose $\hat \rho$ such that, for all $k \in [N]$,
    \begin{align*}
        \mu\left(\left\{
        % (x,y) \in \R^{N \times d_1}\times \R^{N\times N \times d_2}:\\ 
        |\rho(x_k,\sum_{i=1}^N \alpha(x_i, \sum_{j \neq i} \varphi(x_i,y_{ij},x_j))) - \hat\rho(x_k,\sum_{i=1}^N \alpha(x_i, \sum_{j \neq i} \varphi(x_i,y_{ij},x_j)))|_p \geq \frac{\bar \varepsilon}{N}\right\}\right)\leq \frac{\varepsilon}{3N}.
    \end{align*}
    Note that $\hat \rho$ is Lipschitz continuous with some constant $K_{\hat \rho}>0$.
    Then, choose $\hat \alpha$ such that, for all $i \in [N]$,
    \begin{align*}
        \mu\left(\left\{\left|\alpha(x_i,\sum_{j \neq i}\varphi(x_i,y_{ij},x_j)) - \hat \alpha(x_i,\sum_{j \neq i}\varphi(x_i,y_{ij},x_j)) \right|_p \geq \frac{\bar \varepsilon}{N^2K_{\hat \rho}} \right\}\right) \leq \frac{\varepsilon}{3N}.
    \end{align*}
    Finally, $\hat \alpha$ is also Lipschitz continuous with $K_{\hat \alpha}>0$, and we can choose $\hat \varphi$ such that, for all $i,j \in [N]$,
    \begin{align*}
        \mu(\{|\varphi(x_i,y_{ij},x_j) - \hat \varphi(x_i,y_{ij},x_j)|_p \geq \frac{\bar \varepsilon}{ N^2(N-1)K_{\hat \rho}K_{\hat \alpha}} \}) \leq \frac{\varepsilon}{3N(N-1)}.
    \end{align*}
    The choices of $\hat \rho$, $\hat \alpha$, and  $\hat \varphi$ are valid due to Proposition \ref{thm:universality_of_measurable_thomas}.

    Next, we establish several inclusion that will help us to bound the probability.
    \begin{align*}
        &\{|g(x)-\hat g(x)|_p \geq \bar \varepsilon\} = \left\{ \left|
        \begin{array}{l}
              \rho(x_1,\sum_{i=1}^N \alpha(x_i, \sum_{j \neq i} \varphi(x_i,y_{ij},x_j)))\\  - \hat\rho(x_1,\sum_{i=1}^N \hat\alpha(x_i, \sum_{j \neq i} \hat\varphi(x_i,y_{ij},x_j)))\\
              \multicolumn{1}{c}{\vdots}\\ %\vdots\hskip8em\relax
             \rho(x_N,\sum_{i=1}^N \alpha(x_i, \sum_{j \neq i} \varphi(x_i,y_{ij},x_j)))\\ {}- \hat\rho(x_N,\sum_{i=1}^N \hat\alpha(x_i, \sum_{j \neq i} \hat\varphi(x_i,y_{ij},x_j)))
        \end{array}  \right|_p \geq \bar \varepsilon \right\}\ \\
        \subseteq &\bigcup_{k \in [N]} \{|\rho(x_k,\sum_{i=1}^N \alpha(x_i, \sum_{j      \neq i} \varphi(x_i,y_{ij},x_j))) - \hat\rho(x_k,\sum_{i=1}^N      
            \hat\alpha(x_i, \sum_{j \neq i} \hat\varphi(x_i,y_{ij},x_j)))|_p \geq \frac{\bar \varepsilon}{N}\} \\
        \subseteq &\bigcup_{k \in [N]}\bigg(
            \{|\rho(x_k,\sum_{i=1}^N \alpha(x_i, \sum_{j \neq i} \varphi(x_i,y_{ij},x_j))) - \hat\rho(x_k,\sum_{i=1}^N \alpha(x_i, \sum_{j \neq i} \varphi(x_i,y_{ij},x_j)))|_p \geq \frac{\bar \varepsilon}{ N}\}\\
            &\cup \{|\hat\rho(x_k,\sum_{i=1}^N \alpha(x_i, \sum_{j \neq i} \varphi(x_i,y_{ij},x_j))) - \hat\rho(x_k,\sum_{i=1}^N \hat\alpha(x_i, \sum_{j \neq i} \varphi(x_i,y_{ij},x_j)))|_p \geq \frac{\bar \varepsilon}{ N}\}\\
            &\cup \{|\hat\rho(x_k,\sum_{i=1}^N \hat\alpha(x_i, \sum_{j \neq i} \varphi(x_i,y_{ij},x_j))) - \hat\rho(x_k,\sum_{i=1}^N \hat\alpha(x_i, \sum_{j \neq i} \hat\varphi(x_i,y_{ij},x_j)))|_p \geq \frac{\bar \varepsilon}{ N}\}\bigg).
        \end{align*}
        The error on the first set in this expression can be controlled because of the choice of $\hat \rho$. Regarding the second set, we can write
        \begin{align*}
        &\{|\hat\rho(x_k,\sum_{i=1}^N \alpha(x_i, \sum_{j \neq i} \varphi(x_i,y_{ij},x_j))) - \hat\rho(x_k,\sum_{i=1}^N \hat\alpha(x_i, \sum_{j \neq i} \varphi(x_i,y_{ij},x_j)))|_p \geq \frac{\bar \varepsilon}{ N}\}\\
        &\subseteq
            \{|\sum_{i=1}^N \alpha(x_i,\sum_{j \neq i} \varphi(x_i,y_{ij},x_j)) -  \sum_{i=1}^N\hat\alpha(x_i,\sum_{j \neq i} \varphi(x_i,y_{ij},x_j))|_p \geq \frac{\bar \varepsilon}{ NK_{\hat \rho}}\}\\
        &\subseteq \bigcup_{i \in [N]}\{|\alpha(x_i,\sum_{j \neq i} \varphi(x_i,y_{ij},x_j)) -  \hat\alpha(x_i,\sum_{j \neq i} \varphi(x_i,y_{ij},x_j))|_p \geq \frac{\bar \varepsilon}{ N^2K_{\hat \rho}}\}.
        \end{align*}
        Note that this last expression does not depend on $k$ anymore, so there is no need to include it in the union over all $k\in [N]$.
        Similarly, for the third term we find
        \begin{align*}
            &\{|\hat\rho(x_k,\sum_{i=1}^N \hat\alpha(x_i, \sum_{j \neq i} \varphi(x_i,y_{ij},x_j))) - \hat\rho(x_k,\sum_{i=1}^N \hat\alpha(x_i, \sum_{j \neq i} \hat\varphi(x_i,y_{ij},x_j)))|_p \geq \frac{\bar \varepsilon}{ N}\}\\
            &\subseteq\{|\sum_{i=1}^N \hat\alpha(x_i,\sum_{j \neq i} 
            \varphi(x_i,y_{ij},x_j)) -  \sum_{i=1}^N\hat\alpha(x_i,\sum_{j \neq i} \hat\varphi(x_i,y_{ij},x_j))|_p \geq \frac{\bar \varepsilon}{ NK_{\hat \rho}}\}\\
            &\subseteq\bigg(\bigcup_{i \in [N]}\{|\hat\alpha(x_i,\sum_{j \neq i} \varphi(x_i,y_{ij},x_j)) -  \hat\alpha(x_i,\sum_{j \neq i} \hat\varphi(x_i,y_{ij},x_j))|_p \geq \frac{\bar \varepsilon}{ N^2K_{\hat \rho}}\} \bigg)\\
            &\subseteq \bigg(\bigcup_{i \in [N]}\{|\sum_{j \neq i} \varphi(x_i,y_{ij},x_j) -  \sum_{j \neq i} \hat\varphi(x_i,y_{ij},x_j)|_p \geq \frac{\bar \varepsilon}{ N^2K_{\hat \rho}K_{\hat \alpha}}\} \bigg)\\
            &\subseteq \bigg(\bigcup_{i \in [N]}\bigcup_{j \neq i}\{|\varphi(x_i,y_{ij},x_j) -\hat\varphi(x_i,y_{ij},x_j)|_p \geq \frac{\bar \varepsilon}{ N^2(N-1)K_{\hat \rho}K_{\hat \alpha}}\} \bigg).
        \end{align*}

    From this we can conclude that 
    \begin{align*}
        \mu(\{|g(x)-\hat g(x)|_p \geq \bar \varepsilon\}) \leq N\frac{\varepsilon}{3N} + N\frac{\varepsilon}{3N} + N(N-1)\frac{\varepsilon}{3N(N-1)} = \varepsilon.
    \end{align*}
\end{proof}

Due to Proposition \ref{prop:repr_of_GPEqui} and Proposition \ref{prop:universality_of_composition}, we know that PENNs can approximate any permutation-equivariant node-labelling function on $\widetilde{\mathcal{D}}$ arbitrarily well in probability.

Furthermore, if we can approximate such functions on $\widetilde{\mathcal{D}}$, then we can also approximate permutation-equivariant functions on $\mathcal{D}$, as we will show next.

Clearly, any function $\tau: \mathcal{D}\to \R^N$ can be extended to a function $\widetilde{\tau}:\widetilde{\mathcal{D}} \to \R^N$ by 
\begin{align*}
    \widetilde{\tau}(\widetilde a,\ell) = \widetilde{\tau}((a,\bar \sigma),\ell):= \tau(a,\ell).
\end{align*}
Moreover, the following lemma holds.
\begin{lem}
If  $\tau$ is permutation-equivariant, then $\widetilde{\tau}$ is also permutation-equivariant.
\end{lem}
\begin{proof}
    For any permutation $\gamma \in S_N$ and any $(\widetilde a,\ell)= ((a, \bar \sigma),\ell) \in \widetilde{\mathcal{D}}$, it holds that
\begin{align*}
    \widetilde{\tau}(\gamma(\widetilde a,\ell)) &= \widetilde{\tau}(\gamma(\widetilde a), \gamma(\ell)) = \widetilde{\tau}( (\gamma(a), \gamma(\bar \sigma)), \gamma(\ell)) = \tau(\gamma(a),\gamma(\ell))\\ &= \gamma(\tau(a,\ell)) = \gamma(\widetilde{\tau}((a, \bar \sigma),\ell)) = \gamma(\widetilde{\tau}(\widetilde a,\ell)).
\end{align*}
\end{proof}
Then, from Proposition \ref{prop:repr_of_GPEqui}, it immediately follows that we can also find a similar representation for any permutation-equivariant function $\tau$ on $\mathcal{D}$:
\begin{cor}\label{cor:repr_of_tau}
Let $\tau: \mathcal{D}\to \R^N$ be a permutation-equivariant node-labelling function and $\widetilde{\tau}:\widetilde{\mathcal{D}} \to \R^N$ its extension to $\widetilde{\mathcal{D}}$.
Further, for any permutation $\sigma \in S(N)$, let  $f_\sigma:\R^{N\times d} \to \R^{N\times (d+1)}$ be the function that equips node features $a$ with the order $\bar\sigma,$
\begin{align*}
    f_\sigma(a) = \begin{pmatrix}
        a_1, \bar \sigma_1\\ \vdots \\ a_N, \bar \sigma_N
    \end{pmatrix}.
\end{align*}

Then there exist dimensions $V, W \in \N$ and functions $\varphi: \mathbb{R}^{2 (d+1)+d'} \rightarrow \mathbb{R}^{V}$, $\alpha: \mathbb{R}^{d+1+V} \rightarrow \mathbb{R}^W$, and $\rho: \mathbb{R}^{d+1+W} \rightarrow \mathbb{R}$ such that,
for any $\sigma \in S_N$,
\begin{align*}
    \tau(a,\ell)_k = \rho\left( f_\sigma(a)_k, \sum_{i=1}^N \alpha\left(f_\sigma(a)_i, \sum_{j \neq i} \varphi\left(f_\sigma(a)_i, \ell_{i, j}, f_\sigma(a)_j\right)\right)\right).
\end{align*}
In particular, we can choose the functions $\rho$, $\alpha$, and $\varphi$ from Proposition \ref{prop:repr_of_GPEqui} whose composition represents $\widetilde \tau$.
\end{cor}
\begin{proof}
Choose $\rho$, $\alpha$, and $\varphi$ as in Proposition \ref{prop:repr_of_GPEqui}, such that they represent the extension $\widetilde{\tau}$ of $\tau$ to $\widetilde{\mathcal{D}}$.
\end{proof}

The above results imply the following procedure to approximate a node-labelling function $\tau: \mathcal{D}\to \R^N$ on graph data $\{g_1,g_2,\ldots\} \subseteq \mathcal{D}$. 
\begin{enumerate}
    \item Augment the data with random node IDs.
    \item Learn $\widetilde \tau$ with the augmented data by approximating $\rho$, $\alpha$, and $\varphi$ from Proposition \ref{prop:repr_of_GPEqui}.
    \item Use the approximation of $\widetilde \tau$ to obtain an approximation of $\tau$ according to Corollary \ref{cor:repr_of_tau}.
\end{enumerate} 

\begin{rem}
    In the proof of Proposition \ref{prop:repr_of_GPEqui}, we need unique node IDs to identify and distinguish the different nodes solely based on their features. Without unique node IDs, there is no guarantee that a representation as described in Proposition \ref{prop:repr_of_GPEqui} can be found. In some situations, it can even be proved that applying PENNs to data without node IDs cannot work. As an example, we refer to the numerical experiment in Section \ref{section:simple_ex}. However, in other situations -- see for example the numerical experiments in Sections \ref{section:exp_1}, \ref{section:exp_2}, and \ref{section:exp_3} -- PENNs work equally well without node IDs. This can have two reasons. Firstly, if the node features already allow us to distinguish the nodes well enough, adding unique node IDs can be redundant. Secondly, it is possible that a simpler representation of the node-labelling function exists, compared to the one constructed in the proof of Proposition \ref{prop:repr_of_GPEqui}. If this simpler representation exists and does not require node IDs, then PENNs can work well without augmenting the nodes with unique IDs.
\end{rem}

\subsection{Extended permutation-equivariant neural networks}
As we will see in the numerical experiments in the following section, it may occur that important information about a node $k\in [N]$, contained in the expression \begin{align*}
   \hat\alpha(a_k, \sum_{j \neq k} \hat\varphi(a_k, \ell_{kj},a_j))
\end{align*}
gets ``lost'' in the summation
\begin{align*}
    \sum_{i=1}^N \hat\alpha(a_i, \sum_{j \neq i} \hat\varphi\left(a_i, \ell_{ij},a_j\right)).
\end{align*}
Without node numbering, this information might be impossible to recover, and even with node numbering, it can be ``hard'' to recover. For this reason, we propose an extended architecture of PENNs called the extended PENN (XPENN), which also contains a third component that aggregates a node-specific signal for each node, based on its connections from and to all other nodes. Hence, XPENNs maintain all the favourable properties of PENNs and are also able to mimic one iteration of message passing. This architecture is even more expressive, since the aggregation is not only across the neighbours, but also across all other nodes, and it considers both incoming and outgoing edges. In theory, under the assumption of unique node IDs, this additional component does not contain new information about the graph. However, it resolves the issue that certain information may be ``lost'' in the summation, or may be ``difficult'' to recover. In practice, it improves the performance in all numerical experiments we consider; see Section \ref{section:numerical_experiments}.
\begin{defn}\label{def:XPENN}
    With $\hat \varphi$ and $\hat \alpha$ as in Definition \ref{def:PENN}, as well as feedforward neural networks $\hat \psi: \R^{2d + 2d'} \to \R^{d_3}$ and $\hat \rho: \R^{d + d_2 + d_3} \to \R$, for some $d_3 \in \N$, we call the function $f: \R^{N \times d} \times \R^{N \times \N \times d'} \to \R^N$ that is, for $k \in [N]$, defined as
    \begin{align*}
        f(a,\ell)_k =
        \hat\rho\left(a_k, \sum_{i=1}^N \hat\alpha\left(a_i, \sum_{j \neq i} \hat\varphi\left(a_i, \ell_{ij},a_j\right)\right), \sum_{j \neq k} \hat \psi(a_k, \ell_{kj}, \ell_{jk}, a_j)\right)
    \end{align*}
    an \emph{extended permutation-equivariant neural network} (XPENN).
\end{defn}

The following lemma demonstrates that, indeed, XPENNs are an extension of PENNs and include all PENNs.
\begin{lem}\label{lem:xpenn=penn}
    For every PENN $f: \R^{N \times d} \times \R^{N \times \N \times d'} \to \R^N$, we find a corresponding XPENN $f^*: \R^{N \times d} \times \R^{N \times \N \times d'} \to \R^N$, with the same output, i.e., for all $(a,\ell) \in \mathcal{D}$,
    $$
        f^*(a,\ell) = f(a, \ell).
    $$
\end{lem}
\begin{proof}
    Let $f: \R^{N \times d} \times \R^{N \times \N \times d'} \to \R^N$ be a PENN consisting of feedforward neural networks $\hat \rho$, $\hat \alpha$, and $\hat \varphi$. 
    Define $\hat \psi \equiv 0$ and $\hat\rho^*: \R^{d + d_2 + d_3} \to \R$ for all $x_1 \in \R^d, x_2 \in \R^{d_2}, x_3 \in \R^{d_3}$ as
    \begin{align*}
        \hat \rho^*(x_1,x_2,x_3) = \hat \rho(x_1,x_2).
    \end{align*}
    Then, for the XPENN $f^*: \R^{N \times d} \times \R^{N \times \N \times d'} \to \R^N$ given by 
    \begin{align*}
        f^*(a,\ell)_k =
        \hat\rho^*\left(a_k, \sum_{i=1}^N \hat\alpha\left(a_i, \sum_{j \neq i} \hat\varphi\left(a_i, \ell_{ij},a_j\right)\right), \sum_{j \neq k} \hat \psi(a_k, \ell_{kj}, \ell_{jk}, a_j)\right)
    \end{align*}
    it holds for all $(a,\ell) \in \mathcal{D}$
    \begin{align*}
        f^*(a,\ell) = f(a, \ell). 
    \end{align*}
\end{proof}

From Lemma \ref{lem:xpenn=penn}, it follows immediately that XPENNs have the same favourable theoretical properties as PENNs. For example, we can find a corresponding representation of permutation-equivariant node-labelling functions.
\begin{cor}
         A function $\widetilde{\tau}:\widetilde{\mathcal{D}} \to \R^N$ is permutation-equivariant as in Definition \ref{def:permutation_equi} if and only if there exist dimensions $V \leq N^2d'$, $W \leq N(d+1)+ N^2d'$, $W' \in \N$ and functions $\varphi: \R^{2 (d+1)+d'} \rightarrow \R^{V}$, $\alpha: \R^{d+1+V} \rightarrow \R^W$, $\psi: \R^{2(d+1) + 2d'} \to \R^{W'}$, and $\rho: \R^{d+1+W+W'} \rightarrow \mathbb{R}$ such that, for all $(\widetilde{a},\ell) \in \widetilde{\mathcal{D}}$ and for all $k=1, \ldots, N$:
     \begin{align}\label{line:only_PI_on_tilde_D}
         \widetilde{\tau}(\widetilde{a},\ell)_k =\rho\left(\tilde a_k, \sum_{i=1}^N \alpha\left(\tilde a_i, \sum_{j \neq i} \varphi\left(\tilde a_i, \ell_{i, j}, \tilde a_j\right)\right),\sum_{j \neq k} \hat \psi(\tilde a_k, \ell_{kj}, \ell_{jk}, \tilde a_j) \right).
     \end{align}
\end{cor}
\begin{proof}
    This follows from Lemma \ref{lem:xpenn=penn} and Proposition \ref{prop:repr_of_GPEqui}.
\end{proof}

Similarly, we can recover universal approximation for XPENNs.
\begin{cor}
    Let $\rho$, $\alpha$, $\varphi$, and $\psi$ be Borel measurable functions $\varphi: \R^{2d_1+d_2} \to \R^{d_3}$, $\alpha: \R^{d_1+d_3}\to \R^{d_4}$, $\psi: \R^{2d_1 + 2d_2} \to \R^{d_5}$, and $\rho: \R^{d_1+d_4+d_5}\to \R$, for arbitrary dimensions $d_1,\ldots,d_5\in \N$.
    Define $g:\R^{N \times d_1}\times \R^{N\times N \times d_2} \to \R^N$, for all $k \in [N]$, by 
    \begin{align*}
       g(x,y)_k :=
           \rho\left(x_k,\sum_{i=1}^N\alpha(x_i,\sum_{j \neq i} \varphi(x_i, y_{ij},x_j)), \sum_{j \neq k} \psi(x_k, y_{kj}, y_{jk}, x_j)\right).
    \end{align*}
    Then, for any probability measure $\mu$ on $(\R^{N \times d_1}\times \R^{N\times N \times d_2}, \mathcal{B}(\R^{N \times d_1}\times \R^{N\times N \times d_2}))$, and any $\varepsilon,\bar \varepsilon>0$, there exist feedforward neural networks $\hat \rho$, $\hat \alpha$, $\hat \varphi$, and $\hat \psi$ such that, for the function $\hat g:\R^{N \times d_1}\times \R^{N\times N \times d_2} \to \R^N$, which is, for $k \in [N]$, defined by
    \begin{align*}
        \hat g (x,y)_k := 
           \hat\rho\left(x_k,\sum_{i=1}^N\hat\alpha(x_i,\sum_{j \neq i} \hat\varphi(x_i, y_{ij},x_j)), \sum_{j \neq k} \hat\psi(x_k, y_{kj}, y_{jk}, x_j)  \right),
    \end{align*}
    it holds that
    \begin{align*}
        \mu \left( \{(x,y) \in \R^{N \times d_1}\times \R^{N\times N \times d_2}: |g(x,y) - \hat g(x,y)|_p > \bar \varepsilon  \}\right) < \varepsilon.
    \end{align*}
\end{cor}

%###################################################
\section{Numerical experiments}\label{section:numerical_experiments}

In this section, we conduct four numerical experiments. In a first experiment, we consider carefully constructed financial networks, where the optimal amount of bailout capital and the optimal allocation is known. We show that GNNs and XPENNs are able to solve the inner problem, i.e., they find the optimal allocation of the bailout capital. Additionally, we point out some limitations which we observe for FNNs and PENNs.
In our second experiment, we investigate how GNNs, (X)PENNs, FNNs, and some benchmark approaches perform in computing the inner risk, i.e., minimising the risk of loss at some fixed level of bailout capital, on more complex synthetic data, for which the optimal allocation is not known. 
In our third experiment, we approximate the systemic risk by searching for the minimal amount of bailout capital needed to secure the system and compare the performance of GNNs, (X)PENNs, FNNs, and benchmark approaches. 
Finally, in the last experiment, we repeat the third experiment with a non-linear risk measure.

The code to replicate all results of the numerical experiments can be found in a public GitHub repository.\footnote{ https://github.com/NiklasMWeber/ComputingSystemicRiskMeasures.git}

\subsection{Default cascade networks and star-shaped networks}
\label{section:simple_ex}
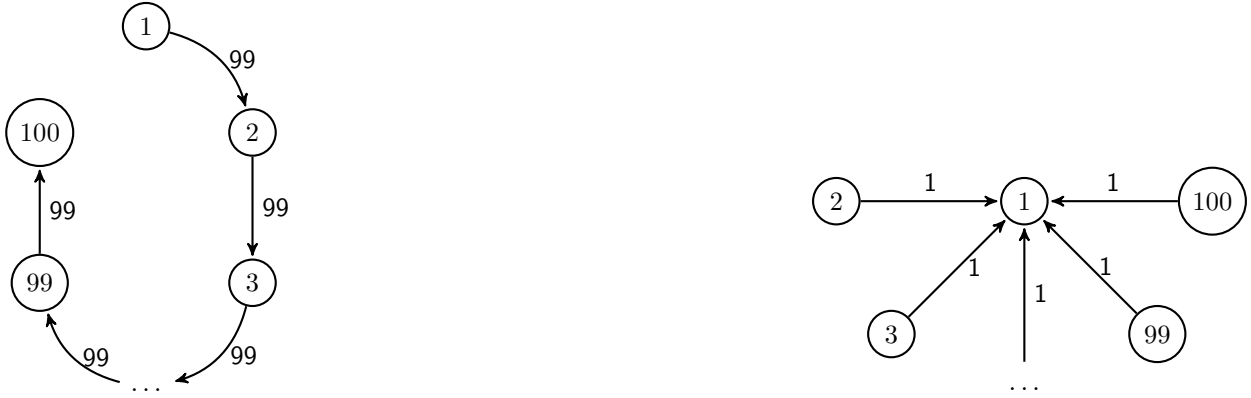
\begin{figure}[h!]
    \centering
\begin{tikzpicture}[->,>=stealth',shorten >=1pt,auto,node distance=2cm, thick, main node/.style={circle,draw,font=\sffamily}, scale=0.8]

  \node[main node] (0) {$1$};
  \node[main node] (1) [below right of=0] {$2$};
  \node[main node] (2) [below of=1] {$3$};
  \node[main node] (3) [below left of=2, draw=none] {$\ldots$}; \node[main node] (4) [above left of=3] {$99$};
  \node[main node] (5) [above  of=4] {$100$};

  \path[every node/.style={font=\sffamily}]
    (0) edge[bend left] node [right] {100} (1)
    (1) edge node [right] {100} (2)
    (2) edge[bend left] node [right] {100} (3)
    (3) edge[bend left] node [right] {100} (4)
    (4) edge node [right] {100} (5);
\end{tikzpicture}
\hfill
\begin{tikzpicture}[->,>=stealth',shorten >=1pt,auto,node distance=2.5cm, thick, main node/.style={circle,draw,font=\sffamily}, scale = 0.8]

  \node[main node] (0) {$1$};
  \node[main node] (1) [left of=0] {$2$};
  \node[main node] (2) [below left of=0] {$3$};
  \node[main node] (3) [below of=0, draw=none] {$\ldots$}; \node[main node] (4) [below right of=0] {$99$};
  \node[main node] (5) [right of=0] {$100$};

  \path[every node/.style={font=\sffamily}]
    (1) edge node [above] {2} (0)
    (2) edge node [right] {2} (0)
    (3) edge node [right] {2} (0)
    (4) edge node [right] {2} (0)
    (5) edge node [above] {2} (0);
\end{tikzpicture}

    \caption{Both network types for $N=100, i=1$. On the left, the default cascade starting at $i=1.$ On the right, the star-shaped network with centre $i=1.$}
    \label{fig:simple_example}
\end{figure}

In this first experiment, we design stylised networks to demonstrate how useful GNNs and XPENNs can be for problems such as learning effective bailout strategies.
\paragraph{Data} We carefully construct four datasets $G_{10}, G_{20}, G_{50}$, and $G_{100} \in \mathcal{G}$ containing realisations of stochastic financial networks of increasing size. Each respective dataset contains realisations with $N\in \{10,20,50,100\}$ nodes. Specifically, for every $i \in 1,\ldots,N$, each dataset contains two distinct realisations of the network $G_N$:  One realisation is a default cascade. This means that every node $j \neq i-1$ has a liability of $N$ units towards the next node $j+1$. Node $N$ will have a liability towards node 1, since there is no node $N+1$. The only link that is missing in the chain is the liability from node $i-1$ towards $i$. Therefore, the network is in fact a cascade of liabilities starting at node $i$ and ending in $i-1$. In the other realisation, each node $j \neq i$ has a liability of two towards $i$, hence forming a star-shaped network with $i$ in the centre. In both cases, all nodes have external assets equal to 1. Figure \ref{fig:simple_example} illustrates both network types.

For each  $N\in \{10, 20, 50, 100\}$,  these $2N$ realisations -- two for every $i \in 1,\ldots,N$ -- create datasets which share one critical feature. We know that they can be fully secured, i.e., there are no defaults, with $N-1$ units of bailout capital. In order to secure a cascade-type network, we need to inject all capital into the starting point of the cascade. On the other hand, to rescue a star-shaped network, we must distribute the bailout capital equally, i.e., allocate one unit to each node, among the nodes with a liability towards the centre node.

\paragraph{Problem} The problem we aim to solve is to find the inner risk
\begin{align*}
    \rho^I_c(G_N) = \min_{Y \in \mathcal{C}^c} \left\{ \eta\left(\Lambda\big((A+Y,L)\big)\right) \right \},
\end{align*}
with allocations in
\begin{align*}
    \mathcal{C}^c = \left\{ Y \in L^\infty(\R_+^N) \,\left|\, \sum_{n=1}^N Y_n =  c\right. \right \},
\end{align*}
% \begin{Notes}{Note}
for $c=N-1$, and $\eta$ a univariate risk measure as in Assumption \ref{assume}.
For the sake of simplicity, we choose $\eta = \myE{}{\cdot}$ to be the expectation and $\Lambda$ the total payment shortfall from Example \ref{exa:lambda:totalpaymentshortfall}.\\
% \end{Notes}

We know that, for each $N$, the theoretical minimal loss is zero if we distribute the bailout $c = N-1$ optimally, as explained above. 
To learn the optimal bailout strategy, 
we deploy a variation of Algorithm \ref{Algo:1} from Section \ref{chapter:comp_theory}, where: 1) we fix the bailout capital and ignore all steps of the probabilistic bisection search, and  2) we directly generate all $2N$ networks. In particular, in this example with finite sample space, there is no need to sample randomly from $(A,L)$. The models that we consider are the following neural network architectures.

\paragraph{Models}
\begin{itemize}[wide]
    \item \textbf{None:} 
    Here, we assume that there is no bailout capital to be distributed. This strategy illustrates the risk of the financial network without any intervention.
    \item \textbf{GNN:} This model is a GNN consisting of two SAGEConv layers (\ref{def:SAGEConv}) with a hidden layer width 10 and sigmoid activation in between. The initial node features are the external assets, which equal 1. This means all nodes start with identical representations, and learning effective message passing is the only possibility for the model to obtain information about the network.
    
    \item\textbf{FNN(L):} This is a vanilla feedforward neural network with two fully connected layers of width $N$, with sigmoid activation in between. Of course, the identical assets alone are not informative input features. To make it a fair comparison between the GNN and a feedforward neural network, we provide the complete liability matrix as input (hence FNN(L)). This neural network theoretically has all information available to perfectly distribute the bailout capital.

    \item\textbf{XPENN:} We also consider the XPENN architecture from the previous section. In fact, this toy experiment showcases well which problems can occur for the standard PENN from Definition \ref{def:PENN}, and why they can be overcome using the XPENN from Definition \ref{def:XPENN}. We discuss the standard PENN at the end of this section. For comparison with GNN and FNN(L), we utilise the XPENN without node IDs and choose single-layer feedforward neural networks of width 10 for $\hat \rho$, $\hat \alpha$, $\hat \varphi$, and a two-layer neural network of width 10 for $\hat \psi$, with sigmoid activation.
\end{itemize}

\paragraph{Task 1 - Overfitting}
\begin{table}[ht]
    \centering
    \begin{tabular}{l r r r r c}
         $N$ & Model & Epoch & Inner Risk & Capital & Time (mm:ss) \\
         \hline
         10 & None & - & 27.00 & 0.00 & - \\
         10 & GNN & 1000 & 0.00 & 9.00 & 01:09 \\
         10 & XPENN & 1000 & 0.00 & 9.00 & 00:47 \\
         10 & FNN(L) & 1000 & 0.23 & 9.00 & 00:23 \\
         \hline
         20 & None & - & 104.50 & 0.00 & - \\
         20 & GNN & 1000 & 0.00 & 19.00 & 02:45 \\
         20 & XPENN & 1000 & 0.00 & 19.00 & 01:52 \\
         20 & FNN(L) & 1000 & 2.68 & 19.00 & 00:59 \\
         \hline
         50 & None & - & 637.00 & 0.00 & - \\
         50 & GNN & 1000 & 0.00 & 49.00 &  08:49 \\
         50 & XPENN & 1000 & 0.00 & 49.00 & 07:42 \\
         50 & FNN(L) & 1000 & 49.78 & 49.00 & 04:04 \\
         \hline
         100 & None & - & 2524.50 & 0.00 & - \\
         100 & GNN & 1000 & 0.00 & 99.00 & 24:02 \\
         100 & XPENN & 1000 & 0.00 & 99.00 & 21:40\\
         100 & FNN(L) & 1000 & 1200.08 & 99.00 & 21:38
    \end{tabular}
    \caption{Inner risk of GNN, XPENN and FNN(L) model on the stylised datasets after training for 1000 epochs.}
    \label{tab:toy_ex_overfit}
\end{table}
% \begin{Notes}{Note}
We perform training runs on all four datasets and investigate if we can teach the models which perfect bailout strategy corresponds to which network structure. If the models are able to learn, this indicates that their complexity is large enough. We optimise the models with suitable learning rates\footnote{In the experiments of Sections \ref{section:simple_ex}, \ref{section:exp_1}, \ref{section:exp_2}, and \ref{section:exp_3}, for all models we considered learning rates in
$\{10^{-2}, 10^{-3}, 10^{-4}, 10^{-5}\}$,
and the best results were selected by grid search.}
using the Adam optimiser, which is a sophisticated version of stochastic gradient descent, and train for 1000 epochs. It is worth pointing out that the first layer of the FNN(L) network connects $N+N^2$ neurons on one side, to $N$ neurons on the other side. Hence, the number of parameters we are training (over $N^3)$ is a lot higher than the number of data points in the training set ($2N$). With this high number of parameters one has to expect that fitting the model to the data should not be a problem, whereas the number of parameters and hence also the time it takes to train them will grow dramatically as we increase $N$. Meanwhile, the GNN's number of parameters is independent from the graph size. Instead, it is driven by the weight matrices of constant size (in our case $2 \times 10$) that map current state and neighbourhood state of each node to its new node state in every step of message passing. As the size of the financial networks increases, more calculations will be involved in each step of message passing. Hence, the computation time will increase as well, but not as dramatically, especially since our networks are sparse. The number of parameters of the XPENN architecture is also independent from the size of the network. However, it calculates a signal for every node pair, even if there is no edge between them. Therefore, it cannot benefit from the fact that the networks are sparse and we expect computation time to increase quadratically.

For $N=10$ (see Table \ref{tab:toy_ex_overfit}) without any bailout capital the average loss of the dataset equals 27 units. The GNN and XPENN are both able to learn the perfect bailout capital allocation within 1000 epochs (actually already earlier), which takes about one minute on our machine.\footnote{
We used a machine with 16 GB of RAM and an AMD Ryzen 5 5600X 6-Core 3.70 GHz CPU.} In $1000$ epochs ($<1$ min), the FNN(L) model learns a strategy that leads to a loss of $0.23$, or roughly 1\% of the loss without bailout.
Doubling the network size to $N=20$, the loss without intervention amounts to $104.5$. GNN and XPENN learn perfect bailout in $1000$ epochs (3 and 2 minutes, respectively) and the FNN(L) arrives at a loss of $2.68$ ($<1\%$) after $1000$ epochs (1 min).
At $N=50$, within 9 minutes the GNN learns perfect bailout, for which the XPENN needs 8 minutes. The FNN(L) takes only 4 minutes, but merely decreases the loss from $637$ without intervention to $49.78$ ($\sim8\%$). Finally, when $N=100$, the parameter space of the FNN(L) is so high-dimensional that the learning rate for training has to be very small ($0.00001$) to wiggle the parameters in the right direction. After $1000$ epochs of training the loss is still very high ($1200$) compared to no intervention ($2524.5$) with no signs of convergence. The GNN and the XPENN handle this case better and reach perfect bailout in under 25 minutes.

The results of this experiment suggest that relatively good, but not perfect, bailout from a feedforward neural network, that takes in the liability matrix of a network is possible when the network size is relatively small. GNN and XPENN, on the other hand, seem to be more promising architectures that always find perfect bailout within 1000 epochs (actually earlier, as we will see in the next experiment). Regarding computation time, we see the following. When roughly doubling $N$ from 10 to 20, from 20 to 50, and from 50 to 100, the runtime of GNN and XPENN, scales roughly with some factor between 2 and 3 (only the XPENN is closer to 4 once, from 20 to 50). The FNN(L) on the other hand seems to scale worse, as we expected. It roughly triples from 10 to 20, it quadruples from 20 to 50, and from 50 to 100 the factor is over 5. It is worth noting, that 
we are not training in batches or mini-batches, but perform a gradient step for each realisation in the dataset individually. Hence, there is most likely a lot of overhead that might hide the effect of the number of parameters on the computing time.  
Nevertheless, the numbers we see are an indicator that the FNN(L) does scale worse than GNN and XPENN. Maybe bigger values of $N$ and training in (mini-)batches would make the expected scaling properties more observable.
% \end{Notes}

\paragraph{Task 2 - Generalisation}
\begin{table}[ht]
    \centering
    \begin{tabular}{l r r r r r c}
         $N$ & Model & Epoch & Train Risk & Test Risk & Capital &Time (mm:ss)\\
         \hline
         10 & None & - & 28.20 & 23.40 & 0.00 & - \\
         10 & GNN & 10 & 0.49 & 0.47 & 9.00 & 00:01 \\
         10 & GNN & 350 & 0.00 & 0.00 & 9.00 & 00:18 \\
         10 & XPENN & 10 & 11.15 & 8.47 & 9.00 & 00:01 \\
         10 & XPENN & 30 & 0.46 &  0.34 & 9.00 & 00:02 \\
         10 & XPENN & 270 & 0.00 & 0.00 & 9.00 & 00:08 \\ 
         10 & FNN(L) & 10 & 11.34 &  10.49 & 9.00 & 00:01 \\
         10 & FNN(L) & 1000 & 0.69 & 12.55 & 9.00 & 00:17 \\
         \hline
         20 & None & - & 104.50 & 104.50 & 0.00 & - \\
         20 & GNN & 10 & 0.07 & 0.07 & 19.00 & 00:01 \\
         20 & GNN & 110 & 0.00 & 0.00 & 19.00 & 00:13 \\
         20 & XPENN & 10 & 38.75 & 38.75 & 19.00 & 00:01 \\
         20 & XPENN & 30 & 0.40 & 0.40 & 19.00 & 00:02 \\
         20 & XPENN & 230 & 0.00 & 0.00 & 19.00 & 00:20 \\
         20 & FNN(L) & 10 & 46.14 &  48.30 & 19.00 & 00:01 \\
         20 & FNN(L) & 1000 & 2.16 & 49.96 & 19.00 & 00:44 \\
         \hline
         50 & None & - & 644.84 & 613.48 & 0.00 & - \\
         50 & GNN & 10 & 325.94 & 308.79 & 49.00 & 00:04 \\
         50 & GNN & 40 & 0.00 & 0.00 & 49.00 & 00:14 \\
         50 & XPENN & 10 & 11.19 & 10.60 & 49.00 & 00:04 \\
         50 & XPENN & 20 & 1.17 & 1.11 & 49.00 & 00:07 \\
         50 & XPENN & 150 & 0.00 & 0.00 & 49.00 & 00:52 \\
         50 & FNN(L) & 10 & 305.02 &  299.80 & 49.00 & 00:01 \\
         50 & FNN(L) & 1000 & 23.10 & 314.88 & 49.00 & 03:28 \\
         \hline
         100 & None & - & 2524.50 & 2524.50 & 0.00 & - \\
         100 & GNN & 10 & 49.65 & 49.65 & 99.00 & 00:09 \\
         100 & GNN & 110 & 0.00 & 0.00 & 99.00 & 01:53 \\
         100 & XPENN & 10 & 7.49 & 7.49 & 99.00 & 00:09 \\
         100 & XPENN & 100 & 0.00 & 0.00 & 99.00 & 01:39 \\
         100 & FNN(L) & 10 & 1227.93 &  1246.98 & 99.00 & 00:11 \\
         100 & FNN(L) & 1000 & 455.16 & 1333.04 & 99.00 & 17:11 \\
    \end{tabular}
    \caption{Inner risk per epoch of GNN, XPENN, and FNN(L) model on the train and test set of the stylised network data. GNN and XPENN achieve better results compared to a FNN(L).}
    \label{tab:toy_ex_generalise}
\end{table}
% \begin{Notes}{Note}
Next, we repeat the experiment in a similar manner, because we want to know whether the models that we train generalise to unseen data. We randomly split the dataset of $2N$ networks into two partitions. A training set containing 75\% of the networks, and a test set with the remaining 25\%. We optimise the models over $1000$ epochs utilising the Adam optimiser. We obtain the following results (Table \ref{tab:toy_ex_generalise}). 

For all network sizes, the FNN(L) is able to decrease the loss on the training set. However, the success on the training set cannot be translated into good bailout capital allocations on the test set. It seems that the FNN(L) is overfitting on the train set and does not recognize the permuted graphs in the test set. It is noticeable that GNN and XPENN are able to learn good bailout very fast (most of the time 30 epochs), always reach perfect bailout in less than 1000 epochs, and can translate perfect bailout on the train set into perfect bailout on the test set. The reason for this is that reordering the nodes does not change the messages that GNN and XPENN receive from the other nodes. This invariance is a major advantage compared to the FNN(L). Based on this experiment, GNN and XPENN seem to be promising candidates for problems of this type. 
% Beyond that, the results do not exhibit any surprises regarding performance or run time compared to what we saw in the previous experiment.
% \end{Notes}

\paragraph{Discussion of PENN types}
During our initial experiments, we observed that, unlike the extended PENN (XPENN), which yields promising results, the normal PENN does not seem to work well, even though in theory it should be able to find perfect bailout. In this paragraph we want to discuss what the problems might be in this specific dataset and how the XPENN can overcome them.

Since all assets are equal, the nodes have identical node features. Hence, it is clear from Definition \ref{def:PENN} that the standard PENN without node IDs will not be able to assign different bailout capital to the different nodes. 
    
However, even with node IDs, the PENN fails to predict good bailout already in the simplest case, $N=10$. We argue that this has two main reasons. On the one hand, the PENN aggregates the \textit{wrong} information in the inner sum $\sum \varphi(\ldots)$. On the other hand, this information is lost in the outer sum over all nodes $\sum_{i=1}^N \alpha(\ldots)$, and cannot be recovered properly by $\rho$. 

What we mean by aggregating the \textit{wrong} information is the following: For any node $i$, the PENN calculates signals based on $L_{ij}$ -- the outgoing edges of $i$ --  while, in this example, it would be more useful to gather information about the incoming edges of $i$. Indeed, the optimal bailout allocation in this dataset is such that 
a node never gets capital if it has incoming liabilities. Information about the outgoing liabilities alone, on the other hand, is not enough to deduce the allocated capital.
This motivates why we consider both directions (incoming and outgoing) in the XPENN architecture (see Definition \ref{def:XPENN}).

To show that aggregating the ``wrong'' data is not the only problem,
we also tested PENNs that consider incoming instead of outgoing liabilities.
However, even then, we saw that the standard PENN is not able to predict good bailout -- and this despite the fact that, in order to predict perfect bailout, it would be enough to recover the sum of incoming liabilities for each node.
Since it is easy to obtain the sum of incoming liabilities of each node from the liability matrix (it is just the column sum of the respective columns), the problem must be obtaining the liability matrix from the outer sum $\sum \alpha(\ldots)$. Additional experiments suggest that these difficulties might arise due to the inhomogeneity of the financial networks, in particular the edge weights. To solve this problem, we designed the XPENN architecture such that it also consists of a node-specific component. As a consequence, graph-wide information given by the outer sum and node-specific information can complement each other, but it is not necessarily required anymore to reconstruct node specific information from the outer sum.

We conclude that, indeed, the data that we generated in this experiment cannot be handled well by the standard PENN, due to the uninformative node features and the inhomogeneity of the financial networks.
The XPENN architecture that we propose manages to overcome these problems by, on the one hand, always considering incoming and outgoing edges, and, on the other hand, considering a node-specific component in which each node can aggregate information about all its connections, both incoming and outgoing.
The next experiments show that, on other types of data, the standard PENN yields very good results as well.

%###################################################
\subsection{Allocation of fixed bailout capital}\label{section:exp_1}

\paragraph{Problem} In this section, we again approximate the inner risk
\begin{align*}
    \rho^I_c(G_N) = \min_{Y \in \mathcal{C}^c} \left\{ \eta\left(\Lambda\big((A+Y,L)\big)\right) \right \},
\end{align*}
with allocations in
\begin{align*}
    \mathcal{C}^c = \left\{ Y \in L^\infty(\R_+^N) \,\left|\, \sum_{n=1}^N Y_n =  c\right. \right \},
\end{align*}
for $c\in \R_+$, and we choose $\eta$ as the expectation and $\Lambda$ as the total payment shortfall from Example \ref{exa:lambda:totalpaymentshortfall}.
However, the stochastic networks we investigate are more complex than before. Therefore, it is not clear which minimal loss could be achieved with the fixed bailout capital that we choose at level $c=50$.

\paragraph{Data} In three case studies, we investigate three different stochastic financial networks that are chosen in a way such that they all contain default cascades which we are seeking to stop. Hence, the networks can be interpreted as formerly healthy financial networks where some kind of shock has already occurred.
\begin{itemize}[wide]
    \item \textbf{Erd\H{o}s--Rényi (ER):} This homogeneous network contains $N=100$ nodes of the same type. Their assets follow independent Beta(2,5)-distributions scaled by a factor of 10. The edges of fixed liability size one between any two nodes are drawn independently with probability $p=0.4$. 
    \item \textbf{Core-Periphery (CP):} This inhomogeneous financial network is an adapted version of networks proposed in \citet{feinstein2017measures}. The network consists of a total of $N = 100$ banks split into two different groups. $T_1 \subset \{1,\ldots,100\}, |T_1|=10,$ contains 10 large banks, and $T_2$ the remaining 90 small banks. A link of a bank in group $i$ to a bank in group $j$, $1\leq i,j\leq2$ is sampled with a probability of $p_{ij}.$ We set $p_{1,1}=0.7$, $p_{1,2}=p_{2,1}=0.3$, $p_{2,2}= 0.1$. If there are links, large banks owe each other 10, small banks owe each other 1, and connections between large and small banks are of size 2. The assets have one-dimensional margins following Beta(2, 5) distributions that are correlated via a Gaussian copula with a correlation of 0.5 between any two banks. Larger banks' assets are scaled by a factor of 50, smaller banks' with a factor of 10.
    \item\textbf{Core--Periphery--fixed (CPf):} This network model is essentially the same as the Core-Periphery model. However, instead of sampling the liabilities new for every realisation of the network, we only sample the liability matrix once, and all realisation share this structure. As a result, we vary only the assets while the network structure is fixed.
\end{itemize}

\paragraph{Models} We utilise various neural network models to learn a distribution $\varphi^\theta$ of the fixed bailout capital (see Section \ref{chapter:comp_theory}), and compare their performance to non-parametric benchmarks, i.e., models that do not require training parameters. All models generate a score for each node. We use softmax activation to transform these scores into a distribution. 
\begin{itemize}[wide]
    \item \textbf{None:} 
    We assume that there is no bailout capital to be distributed. This strategy illustrates the risk of the financial network without any intervention.
    \item\textbf{GNN:} A graph neural network with five SAGEConv layers of message passing. The 10-dimensional initial state of each node consists of assets, incoming liabilities, outgoing liabilities and seven zeros. We use the sigmoid activation function between layers. The hidden states are of dimension 10 and the output layer yields a single score for each node.
    \item\textbf{PENN:} We apply a PENN with single-layer neural networks $\hat \varphi$, $\hat \alpha$, that produce 10-dimensional representations, and a  three-layer neural network $\hat \rho$, with hidden states of dimension 20 producing a score, and ReLU activation. For each node, the initial node features consist of assets, incoming liabilities, and outgoing liabilities. For the performance it did not matter, whether we included node IDs or not.
    \item\textbf{XPENN:} This model is the extended version of the PENN. We use the same neural networks for $\hat \varphi$, $\hat \alpha$, and $\hat \rho$. Additionally, $\hat \psi$ is a two-layer neural network of width 10 and ReLU activation, that provides a 10-dimensional neighbourhood signal for each node.
    \item\textbf{FNN:} This model is a feedforward neural network with three fully connected layers of width 100, and ReLU activation function. As input, it takes a concatenated vector of assets, incoming and outgoing liabilities of all 100 nodes. The output is a score vector of the 100 nodes.

    This approach is inspired by \citet{feng2022deep}. They model a financial network as a random vector $X \in \mathcal{L}(\R^N)$ and obtain the optimal bailout from a neural network $\varphi: X \mapsto \varphi(X).$
    However, in their setting, there is no explicitly modelled interaction between the components of $X$. Rather, the random variable $X$ represents the network after interaction, which they incorporate by allowing correlation between the components of $X$. In this setting, it makes sense to use the realisation of this random variable as representation of the system's state. Applied to our setting, one might think that the realisation of the assets $A$ comes closest to the system state vector $X$. However, this is not entirely true, because $A$ represents some value before the interaction. This interaction depends on the realisation of the liability matrix, which is not captured by the FNN at all. Therefore, this approach can only work well if the dependence of the optimal bailout allocation on the liability matrix is negligible -- which is not true in general. To increase the chances of this approach working, we provide assets, incoming liabilities and outgoing liabilities of each node, instead of only the assets. 
    
    \item\textbf{FNN(L):} This model is similar to the FNN model, but as input, it takes not only assets, incoming and outgoing liabilities of all 100 nodes, but also the liability matrix as one long vector. It can be considered a na\"ive extension of the approach in \cite{feng2022deep} to our setting of stochastic liability matrices. We consider two fully connected layers with width 100.
    \item\textbf{Linear:} In this approach, we fit a linear model that predicts a node's score based on the node's assets, nominal incoming and outgoing liabilities, as well as a bias term.
    \item\textbf{Constant:} Furthermore, we are interested in how much performance we gain by making the allocation random, i.e., scenario-dependent. Therefore, for comparison, we include a benchmark that learns a constant bailout allocation.
    \item\textbf{Uniform:} This is a non-parametric benchmark, where every node gets the same share of bailout capital.
    \item\textbf{Default:} This is a non-parametric benchmark, where every node that defaults without additional bailout capital receives the same amount of bailout capital.
    \item\textbf{Level-1:} This is a non-parametric benchmark where every node whose nominal balance is negative ($A_i-L_{\text{out},i}+ L_{\text{in},i} < 0$) receives the same amount of bailout capital. These nodes are the ``first-round defaults'' and, theoretically, there cannot be additional defaults if all ``first round'' defaults are prevented.
\end{itemize}

\begin{table}[h!]
    \centering
    \begin{tabular}{l|l|l|l}
         Model & Val. Risk $\pm$ std & Test Risk $\pm$ std & Capital  \\
         \hline
         None &261.64 $\pm$ 0.00 &262.08 $\pm$ 0.00&0.00 \\
         Uniform 	&  221.22 $\pm$ 0.00 	&  221.71 $\pm$ 0.00 	&50.00\\
         Default 	&  186.37 $\pm$ 0.00	&  186.96 $\pm$	0.00 &50.00\\
         Level-1 	&  174.32 $\pm$	0.00 &  174.78 $\pm$ 0.00	&50.00\\
         GNN&   173.71 $\pm$	0.01&  174.23 $\pm$	0.01 &50.00\\
         PENN	&  173.71 $\pm$	0.01 & 174.24 $\pm$ 0.01&50.00\\
         XPENN&   172.74 $\pm$ 0.99	&  173.26 $\pm$	1.01 &50.00\\
         FNN &   221.22 $\pm$ 0.01	&  221.71 $\pm$ 0.01	&50.00\\
         FNN(L)  	&  221.22 $\pm$	0.00 &  221.72 $\pm$ 0.01 &50.00\\ 
         Linear  	&  178.66 $\pm$	0.00 &  179.21 $\pm$ 0.00 &50.00\\ 
         Constant  	& 221.24 $\pm$	0.01 &  221.72 $\pm$ 0.01	&50.00 
    \end{tabular}
    \caption{Approximated inner risk and its standard deviation on the validation and test sets, obtained from the bailout allocations learned by the different models on data sampled from the ER network type.}
    \label{tab:my_label}
% \end{table}

% \begin{table}[ht]
\bigskip
    \centering
    \begin{tabular}{l|l|l|l}
         Model & Val. Risk $\pm$ std & Test Risk $\pm$ std & Capital  \\
         \hline
         None 	&  235.56 $\pm$ 0.00 	&  238.34 $\pm$ 0.00 	&0.00 \\
         Uniform 	&  201.45 $\pm$ 0.00	&  203.57 $\pm$	0.00&50.00\\ 
         Default 	&  166.61 $\pm$	0.00&  168.64 $\pm$	0.00 &50.00\\ 		
         Level-1  	&  158.04 $\pm$	0.00 &  160.03 $\pm$ 0.00&  50.00\\	
         GNN 	&  153.15 $\pm$	0.05&  155.17 $\pm$	0.07 &50.00\\ 	
         PENN	&  153.05 $\pm$ 0.06	&  155.09 $\pm$	0.07 &50.00\\
         XPENN 	&  150.21 $\pm$	0.61&  152.14 $\pm$	0.66 &50.00\\ 
         FNN  	&  201.07 $\pm$ 0.48	&  203.20 $\pm$	0.51 &50.00\\ 
         FNN(L)  	&  201.13 $\pm$	0.31&  203.22 $\pm$ 0.35	&50.00\\ 
         Linear  	&  159.80 $\pm$	0.00 &  161.84 $\pm$ 0.01	& 50.00\\ 
         Constant  	&  201.46 $\pm$ 0.01	&  203.59 $\pm$ 0.01	&50.00
    \end{tabular}
    \caption{Approximated inner risk and its standard deviation on the validation and test sets, obtained from the bailout allocations learned by the different models on data sampled from the CP network type.}
    \label{tab:my_label2}
% \end{table}

% \begin{table}[ht]
\bigskip
    \centering
    \begin{tabular}{l|l|l|l}
         Model & Val. Risk $\pm$ std & Test Risk $\pm$ std & Capital  \\
         \hline
         None 	&  296.30 $\pm$	0.00 &  298.50 $\pm$ 0.00	& 0.00 \\
         Uniform 	&  253.98 $\pm$ 0.00	&  255.73 $\pm$	0.00 &50.00\\ 		
         Default 	&  218.20 $\pm$ 0.00	&  219.96 $\pm$ 0.00 &50.00\\
         Level-1  	&  208.57 $\pm$ 0.00	&  210.36 $\pm$	0.00 &50.00\\ 			
         GNN  	&  203.01 $\pm$	2.50 &  204.85 $\pm$ 2.53	&50.00\\ 
         PENN 	&  204.12 $\pm$	3.24 &  205.92 $\pm$ 3.33 &50.00\\
         XPENN 	&  203.57 $\pm$	3.73&  205.40 $\pm$	3.78 &50.00\\ 
         FNN  	&  203.76 $\pm$	9.85&  205.59 $\pm$	9.80 &50.00\\
         FNN(L)  	&  204.37 $\pm$	10.82&  206.18 $\pm$ 10.75	&50.00\\
         Linear  	& 210.40 $\pm$ 0.83	&  212.35 $\pm$ 0.84	&50.00\\ 
         Constant  	&  204.37 $\pm$ 0.01	&  206.27 $\pm$	0.01 &50.00 
    \end{tabular}
    \caption{Approximated inner risk and its standard deviation on the validation and test sets, obtained from the bailout allocations learned by the different models on data sampled from the CPf network type.}
    \label{tab:my_label3}
\end{table}

\paragraph{Results}
The results of all models can be found in Tables \ref{tab:my_label}, \ref{tab:my_label2}, and \ref{tab:my_label3} for the ER, CP, and CPf networks, respectively. 
% \begin{Notes}{Note}
The training algorithm employed is the one presented in Section \ref{chapter:comp_theory}, Algorithm \ref{Algo:1}, with the only variation that we fix the bailout capital and skip all steps that deal with the probabilistic bisection search. Since, in this algorithm, the training data is sampled new continuously in every epoch, we additionally sample two validation and test sets, where we can track and compare the performance of the various models. Furthermore, all models are trained using five different seeds and we report the average and standard deviation of the results.
% \end{Notes}

In the homogeneous ER network (see Table \ref{tab:my_label}), it seems the best model is XPENN. It performs slightly better than GNN and PENN, which essentially reach the same risk level. The higher standard deviation of XPENN is caused by the fact that it sometimes outperforms GNN and PENN, and sometimes performs on the same level (but never worse). The Constant bailout strategy, FNN, and FNN(L), do not work well, and cannot provide bailout allocations that are superior to allocating all capital uniformly. 
% The FNN(L) seems to learn something on the training set, but it is the only model that does not translate learned success to unseen data, a pattern that we also observed in the previous toy experiment. 
The Linear model performs better than the Default method and is not far off the Level-1 method. Surprisingly, the Level-1 method performs strongly, considering it does not require any learning, and is indeed the fourth-best model with a performance similar to GNN and PENN.

In the CP network (see Table \ref{tab:my_label2}), the picture is similar. GNN and PENN are the second-best models only outperformed by XPENN, which again has a higher standard deviation due to the fact that it sometimes finds additional performance and sometimes performs at the same level as GNN and PENN. The Constant, FNN, and FNN(L) models all fail to provide better bailout than Uniform bailout. Level-1 bailout comes fourth -- this time with a bigger gap to GNN and PENN than before -- and closely followed by the Linear model, which again performs better than the Default method and remains close to Level-1.

Finally, in the CPf network (see Table \ref{tab:my_label3}), we observe a change in order. All the more complex ML models (GNN, PENN, XPENN, FNN, FNN(L)) obtain similar results, with GNN being the best and FNN(L) slightly weaker than the rest. Standard deviations are high compared to the other network types, which makes it even more impressive that the Constant bailout is close to the ML models whilst exhibiting small standard deviation. Finally, Level-1, Linear, Default and Uniform follow in this order.

The difference between Linear model and GNN, PENN, or XPENN indicates that, in all three network types, there is structural information that is needed to provide the best bailout. The fact that the FNN(L) does not work in the ER and CP network -- i.e., the cases where there is structural information in the liability matrix -- hints that such structural information cannot easily be captured by feeding the liability matrix into a feedforward neural network. Moreover, the difference between Constant bailout and the respective best model shows how much performance can be gained in the ER and CP networks by making the allocation random. 

The good performance of FNN and FNN(L) in the CPf network shows that these approaches can work if the dependence of the system on the liability matrix is negligible -- in the sense that the influence is constant and hence learnable. This aligns well with the results in \cite{feng2022deep}.

Since, in the CPf network, the liability matrix is constant, this means not only that the interaction between the banks is fixed, it also means that - unlike in the CP network - all banks are always in the same position. This might be the reason why the Constant model actually works in this network, while it performs as poorly as Uniform bailout in the other networks.

%###################################################
\subsection{Computation of systemic risk measures}\label{section:exp_2}
\textbf{Problem} In this section, instead of fixing an arbitrary amount of available bailout capital, we fix a level of risk that we consider acceptable ($b = 100)$ and, choosing $\eta$ as the expectation, search for the systemic risk
\begin{align*}
    \rho_b(G) &= \min_{ Y \in \mathcal{C}} \left \{\left.\sum_{n=1}^N Y_n \,\right|\, \eta \left(\Lambda(A+Y,L)\right) \leq b \right\}
\end{align*}
where
\begin{align*}
    \mathcal{C} = \left\{ Y \in L^\infty(\R_+^N) \,\left|\, \sum_{n=1}^N Y^n \in \R \right. \right \}.
\end{align*}
We study the same \textbf{Data} and \textbf{Models} as before.
The training algorithm employed is the one presented in Section \ref{chapter:comp_theory}; see Algorithm \ref{Algo:1}. As before, since in this algorithm the training data is sampled new continuously in every iteration, we additionally sample two validation and test sets, on which we can track and compare the performance of the various models. All models are trained using five different seeds and we report the average and standard deviation of the results.

\begin{table}[h!]
    \centering
    \begin{tabular}{l|r|r|r}
         Model & Val. Risk $\pm$ std & Test Risk $\pm$ std & Capital $\pm$ std \\
         \hline
         None &261.64 $\pm$ 0.00 &262.08 $\pm$ 0.00&0.00 $\pm$ 0.00 \\
Uniform	 	&  99.49 $\pm$ 0.16	&  99.86 $\pm$ 0.15	&   302.91 $\pm$  0.50\\
Default	 	&  99.35 $\pm$ 0.18	&   99.81 $\pm$	0.18&   147.61 $\pm$ 0.29\\ 	
Level-1	 	&  99.44 $\pm$	0.10&  99.69 $\pm$ 0.10	&   108.54 $\pm$ 0.11\\
GNN		&  99.18 $\pm$	0.15&  	99.78 $\pm$	0.16&   96.03 $\pm$ 0.09 \\ 
PENN	 	&  99.16 $\pm$ 0.27	&   99.74 $\pm$ 0.28	&   96.02 $\pm$ 0.18\\ 
XPENN	& 98.92 $\pm$ 0.46	&   99.51 $\pm$	0.46&   95.83 $\pm$ 0.30\\ 
FNN	 	&  99.50 $\pm$	0.17&  99.85 $\pm$ 0.20	&  303.08 $\pm$ 0.57\\ 
FNN(L)	 	&  99.47 $\pm$ 0.18	&  99.83 $\pm$ 0.20 &   303.24 $\pm$ 0.50\\ 
Linear	 	&  99.34 $\pm$ 0.10	&   99.96 $\pm$	0.09&  	113.77 $\pm$ 0.08\\
Constant	 	&  99.40 $\pm$ 0.18	&  99.76 $\pm$	0.18&   303.26 $\pm$ 0.57 
    \end{tabular}
    \caption{The \textit{Capital} column shows, for each model, the best approximation of the systemic risk, i.e., the smallest amount of bailout capital needed to make the ER network acceptable. The other columns show the inner risk on the validation and test sets.}
    \label{tab:my_label11}
% \end{table}
% \begin{table}[ht]
\bigskip
    \centering
    \begin{tabular}{l|r|r|r}
         Model & Val. Risk $\pm$ std & Test Risk $\pm$ std & Capital $\pm$ std  \\
         \hline
        None 	&  235.56 $\pm$ 0.00 	&  238.34 $\pm$ 0.00 	&0.00 $\pm$ 0.00 \\
Uniform	 	&  99.44 $\pm$ 0.23	&  100.54 $\pm$ 0.23	&  303.14 $\pm$ 0.88\\ 
Default	 	& 99.49 $\pm$ 0.25	&  101.15 $\pm$ 0.25	&   134.11 $\pm$ 0.44\\
Level-1	 	&  99.24 $\pm$ 0.36	& 100.88 $\pm$ 0.36	&   107.55 $\pm$ 0.47\\ 	
GNN	 	& 98.92 $\pm$ 0.29	&  100.41 $\pm$ 0.30	&   87.47 $\pm$ 0.23\\ 
PENN   	&  99.11 $\pm$ 0.36	&  100.66 $\pm$ 0.39	&   87.05 $\pm$ 0.23\\
XPENN	 	&  98.54 $\pm$ 0.55	& 100.01 $\pm$	0.54&   86.67 $\pm$ 0.76\\ 	
FNN	 	&  98.84 $\pm$ 0.33	&  99.91 $\pm$ 0.35	&  303.52 $\pm$ 0.57\\
FNN(L)		&  98.94 $\pm$	0.60& 100.04 $\pm$	0.61&   303.25 $\pm$ 0.98\\
Linear	 	& 98.73 $\pm$ 0.41	&  100.95 $\pm$	 0.37&   113.10 $\pm$ 0.38\\
Constant		&  99.30 $\pm$ 0.18	&  100.40 $\pm$	0.18&   303.74 $\pm$ 0.70
\end{tabular}
    \caption{The \textit{Capital} column shows, for each model, the best approximation of the systemic risk, i.e., the smallest amount of bailout capital needed to make the CP network acceptable. The other columns show the inner risk on the validation and test sets.}
    \label{tab:my_label22}
% \end{table}
% \begin{table}[ht]
\bigskip
    \centering
    \begin{tabular}{l|r|r|r}
         Model & Val. Risk $\pm$ std& Test Risk $\pm$ std & Capital $\pm$ std  \\
         \hline
         None 	&  296.30 $\pm$	0.00 &  298.50 $\pm$ 0.00	& 0.00 $\pm$ 0.00\\
Uniform		&  100.07 $\pm$	0.12&  100.34 $\pm$	0.13&   447.88 $\pm$ 0.70\\ 	
Default	 	& 100.25 $\pm$	0.30&  	101.11 $\pm$ 0.31	&   220.39 $\pm$ 0.85\\ 
Level-1	 	&  99.83 $\pm$ 0.16	&  100.88 $\pm$	0.16&   166.03 $\pm$ 0.31\\
GNN	 	& 99.64 $\pm$ 0.35	&  101.08 $\pm$ 0.36	&   115.72 $\pm$ 0.43\\
PENN  	&  99.74 $\pm$ 0.43	&  101.17 $\pm$	0.44&   	115.09 $\pm$ 0.15\\ 
XPENN 	&  99.82 $\pm$ 0.62	&  101.26 $\pm$	0.63&  114.72 $\pm$ 0.62\\ 
FNN		&  99.57 $\pm$ 0.94	&  100.85 $\pm$ 0.94	&   	116.55 $\pm$ 0.46\\
FNN(L)		&  100.21 $\pm$ 0.29	&  101.48 $\pm$	0.31&   116.80 $\pm$ 0.52\\
Linear	 	&  100.25 $\pm$ 0.30	&  101.37 $\pm$ 0.30	&   139.11 $\pm$ 0.27\\ 
Constant	 	&  99.80 $\pm$ 1.28	&  101.22 $\pm$	1.29&  121.68 $\pm$ 1.01
    \end{tabular}
    \caption{The \textit{Capital} column shows, for each model, the best approximation of the systemic risk, i.e., the smallest amount of bailout capital needed to make the CPf network acceptable. The other columns show the inner risk on the validation and test sets.}
    \label{tab:my_label33}
\end{table}

\paragraph{Results}
The results obtained during the search for the minimal bailout capital in the ER, CP, and CPf networks can be found in Tables \ref{tab:my_label11}, \ref{tab:my_label22}, and \ref{tab:my_label33}. These results are in line with those of the previous experiment, in the sense that the models that obtained the smallest inner risk for a fixed amount of bailout capital require the smallest amount of bailout capital to make the inner risk acceptable with respect to a given risk threshold.

In the ER network (see Table \ref{tab:my_label11}), XPENN is the best model, as it requires the least bailout capital to reduce the inner risk on validation and test set to an acceptable level. GNN and PENN reach similar risk with slightly more capital. The Level-1, Linear, and Default model already require significantly more capital to reach acceptable risk. FNN, FNN(L), Constant, and Uniform require considerably more capital to render the risk acceptable.

In the CP network (Table \ref{tab:my_label22}), XPENN is the strongest model, closely followed by PENN and GNN. With a slightly larger gap, the next best models are Level-1, Linear, and the Default model. FNN, FNN(L), Constant, and Uniform require too much capital to compete with the other models.

Comparing Level-1 or Linear with the Constant model in both, the ER and CP network, we observe that a significant amount of bailout capital can be saved by allowing for scenario-dependent allocations, as opposed to committing to a constant capital allocation. The difference between XPENN, PENN or GNN on the one hand, and Linear or Level-1 model, on the other, illustrates how much bailout capital can be saved by using an approach that processes the liability matrix effectively. The results of FNN and FNN(L) clearly show that these approaches are not applicable to networks with stochastic liability matrices.

Finally, in the CPf network (Table \ref{tab:my_label33}), we see that XPENN, GNN and PENN provide the best results, but also FNN, FNN(L), and even -- although with some gap -- Constant, work well and outperform Linear, Level-1, Default, and Uniform models.
These results underline again that, in the case of a fixed network structure, it is not necessary to utilise models that process the liability matrix effectively. Since the effect is always the same, it can be learned implicitly without being given as input to the model. It is interesting that XPENN, GNN, and PENN still provide the best results. Apparently, it does not harm their performance to receive structural information as input, despite it being constant. 
One explanation why GNN, PENN, and XPENN appear to be slightly stronger than FNN and FNN(L) might be that they obtain good bailout more quickly than FNN(L) and FNN, which makes them more suitable for the training procedure that alternates between updating the bailout capital and learning how to allocate it. This explanation is supported by the observation in the first experiment (Section \ref{section:exp_1}), where GNN and XPENN reached good bailout much more quickly than FNN(L).

The difference between Constant bailout and the best models in the CPf network suggests that, in networks with fixed liability matrices, the benefit of stochastic allocations is smaller than in stochastic networks. This makes sense, since it seems easier to predict whether an institution will need capital or not if there is no uncertainty about its connections from and to other institutions in the network.

The poor performance of Linear and Level-1 in the CPf network is surprising, given their strong performance in the ER and CP networks. This suggests that the optimal bailout of the CPf network, with its randomly fixed liability matrix, does not appear to be driven by a linear combination of assets, incoming and outgoing liabilities, or the ``first-round'' defaults. Apparently, there are more complex interactions at work that can be learnt by the other trainable models.

% ###################################### Entropic ##############
% \begin{Notes}{Note}
\subsection{Systemic risk based on entropic risk measure}\label{section:exp_3}
\paragraph{Problem}
In this section, 
we repeat the experiment of Section \ref{section:exp_2} with the entropic risk measure $\eta:L^\infty \to \R,$
\begin{align*}
    \eta(X) = \frac{1}{\gamma}\log\left(\myE{}{e^{\gamma X}} \right),
\end{align*}
instead of the expectation. This is an example of a risk measure that we can use in the $L^\infty$ setting, as mentioned in Remark \ref{rem:L^1vsL^infty}. Note that, since in our framework the input random variable $X$ models a loss, we consider $X$ directly and not $-X$, as is common in the literature. We choose the risk aversion parameter $\gamma = 0.01$ and leave everything else unchanged.

\paragraph{Results}
The results can be found in Tables \ref{tab:entropic_er}, \ref{tab:entropic_cp}, and \ref{tab:entropic_cpf}. They confirm the findings of the previous experiment. In the ER and CP networks, XPENN, PENN, and GNN are the best models, with XPENN maintaining a slight advantage compared to the others. The permutation-equivariant benchmarks Level-1, Linear, and Default work, but cannot provide as effective capital allocations. FNN, FNN(L), Constant, and Uniform do not yield satisfactory results. In the CPf network, the results suggest again, that for a fixed interbank structure -- besides PENN, XPENN, and GNN -- the models FNN, FNN(L), and even Constant are useful, with FNN and FNN(L) even outperforming GNN.
% \end{Notes}

\begin{table}[h!]
    \centering
    \begin{tabular}{l|r|r|r}
    Model & Val. Risk $\pm$ std & Test Risk $\pm$ std & Capital $\pm$ std \\
    \hline
     None     & 261.64 $\pm$ 0.00 & 262.08 $\pm$ 0.00 & 0.00 $\pm$ 0.00\\
    Uniform  & 99.51 $\pm$ 0.20 & 99.91 $\pm$ 0.20 & 311.42 $\pm$ 0.64 \\
    Default  & 99.32 $\pm$ 0.10 & 99.76 $\pm$ 0.10 & 153.32 $\pm$ 0.16 \\
    Level-1  & 99.45 $\pm$ 0.11 & 99.74 $\pm$ 0.11 & 112.28 $\pm$ 0.12 \\
    GNN      & 98.90 $\pm$ 0.17 & 99.64 $\pm$ 0.18 & 99.77 $\pm$ 0.11 \\
    PENN     & 99.06 $\pm$ 0.21 & 99.78 $\pm$ 0.20 & 99.69 $\pm$ 0.14 \\
    XPENN    & 98.69 $\pm$ 0.55 & 99.43 $\pm$ 0.56 & 99.51 $\pm$ 0.25 \\
    FNN       & 99.50 $\pm$ 0.20 & 99.90 $\pm$ 0.25 & 311.75 $\pm$ 0.75 \\
    FNN(L)      & 99.45 $\pm$ 0.16 & 99.84 $\pm$ 0.20 & 311.93 $\pm$ 0.53 \\
    Linear   & 99.29 $\pm$ 0.14 & 100.08 $\pm$ 0.13 & 117.99 $\pm$ 0.14\\
    Constant & 99.46 $\pm$ 0.11 & 99.86 $\pm$ 0.11 & 311.63 $\pm$ 0.35 \\
\end{tabular}
    \caption{The \textit{Capital} column shows for each model the best approximation of the entropic-risk-measure-based systemic risk, i.e., the smallest amount of bailout capital needed to make the ER network acceptable. The other columns show the inner risk on validation and test sets.}
    \label{tab:entropic_er}
% \end{table}
% \end{table}
%----
% \begin{table}[ht]
\bigskip
% \begin{table}[h!]
    \centering
    \begin{tabular}{l|r|r|r}
    Model & Val. Risk $\pm$ std & Test Risk $\pm$ std & Capital $\pm$ std \\
    \hline
    None     & 235.56 $\pm$ 0.00 & 238.34 $\pm$ 0.00 & 0.00 $\pm$ 0.00 \\
    Uniform  & 99.42 $\pm$ 0.22 & 100.69 $\pm$ 0.23 & 356.30 $\pm$ 0.84 \\
    Default  & 99.83 $\pm$ 0.45 & 103.06 $\pm$ 0.46 & 183.10 $\pm$ 0.80 \\
    Level-1  & 99.00 $\pm$ 0.57 & 101.83 $\pm$ 0.59 & 138.91 $\pm$ 0.72 \\
    GNN      & 98.50 $\pm$ 0.20 & 102.15 $\pm$ 0.24 & 113.97 $\pm$ 0.20 \\
    PENN     & 98.80 $\pm$ 0.49 & 102.51 $\pm$ 0.53 & 113.63 $\pm$ 0.30 \\
    XPENN    & 97.58 $\pm$ 1.25 & 100.99 $\pm$ 1.37 & 112.92 $\pm$ 0.24 \\
    FNN       & 99.16 $\pm$ 0.37 & 100.41 $\pm$ 0.43 & 356.70 $\pm$ 1.20 \\
    FNN(L)      & 99.16 $\pm$ 0.37 & 100.45 $\pm$ 0.36 & 356.55 $\pm$ 0.92 \\
    Linear   & 98.89 $\pm$ 0.45 & 101.06 $\pm$ 0.48 & 149.08 $\pm$ 0.65 \\
    Constant & 99.24 $\pm$ 0.34 & 100.52 $\pm$ 0.34 & 357.00 $\pm$ 1.29 \\
\end{tabular}
    \caption{The \textit{Capital} column shows for each model the best approximation of the entropic-risk-measure-based systemic risk, i.e., the smallest amount of bailout capital needed to make the CP network acceptable. The other columns show the inner risk on validation and test sets.}
    \label{tab:entropic_cp}
% \end{table}
% \end{table}
\bigskip
% \begin{table}[ht]
    \centering
    \begin{tabular}{l|r|r|r}
    Model & Val. Risk $\pm$ std & Test Risk $\pm$ std & Capital $\pm$ std \\
    \hline
    None 	&  296.30 $\pm$	0.00 &  298.50 $\pm$ 0.00	& 0.00 $\pm$ 0.00\\
    Uniform  & 100.27 $\pm$ 0.29 & 100.75 $\pm$ 0.29 & 478.88 $\pm$ 1.52 \\
    Default  & 100.46 $\pm$ 0.42 & 102.01 $\pm$ 0.43 & 267.54 $\pm$ 1.11 \\
    Level-1  & 99.39 $\pm$ 1.17 & 101.28 $\pm$ 1.20 & 196.21 $\pm$ 2.15 \\
    GNN      & 99.97 $\pm$ 2.24 & 103.36 $\pm$ 2.26 & 142.19 $\pm$ 2.74 \\
    PENN     & 98.92 $\pm$ 0.43 & 102.28 $\pm$ 0.48 & 140.89 $\pm$ 0.67 \\
    XPENN    & 98.85 $\pm$ 0.77 & 102.18 $\pm$ 0.77 & 140.90 $\pm$ 0.39 \\
    FNN       & 98.56 $\pm$ 0.38 & 101.53 $\pm$ 0.48 & 140.66 $\pm$ 0.13 \\
    FNN(L)      & 100.52 $\pm$ 3.34 & 103.59 $\pm$ 3.56 & 140.79 $\pm$ 0.25 \\
    Linear   & 100.22 $\pm$ 0.33 & 102.57 $\pm$ 0.33 & 167.06 $\pm$ 0.39 \\
    Constant & 99.00 $\pm$ 0.60 & 102.22 $\pm$ 0.62 & 147.46 $\pm$ 0.39 \\
\end{tabular}
    \caption{The \textit{Capital} column shows for each model the best approximation of the entropic-risk-measure-based systemic risk, i.e., the smallest amount of bailout capital needed to make the CPf network acceptable. The other columns show the inner risk on validation and test sets.}
    \label{tab:entropic_cpf}
\end{table}

%###################################################
\section{Conclusion}
In this work, we extend the notion of systemic risk measures with random allocations under the ``first allocate then aggregate'' paradigm from vector-valued risk factors to network-valued risk factors that are represented by a stochastic asset vector and a stochastic liability matrix.  Under reasonable assumptions on the stochastic financial network, the risk measure, and the aggregation function -- in particular, those that include aggregation functions based on the market-clearing mechanism of \citet{EisNoe} -- we investigate the theoretical properties of these systemic risk measures. We show the existence of optimal random allocations and provide a reformulation of the systemic risk that allows us to derive an iterative optimisation algorithm.

Furthermore, we connect the domain of financial networks to the domain of weighted, directed graphs and -- motivated by this connection -- investigate neural network architectures that are permutation-equivariant. We study \emph{permutation-equivariant neural networks} (PENNs) \cite{herzig2018mapping} and introduce \emph{extended permutation-equivariant neural networks} (XPENNs), which are also permutation-equivariant. For both architectures, we prove universal approximation results, in the sense that they can approximate any permutation-equivariant node-labelling function arbitrarily well in probability.

In numerical experiments with synthetic financial networks, we show that, in order to compute systemic risk measures of networks with stochastic liability matrices, it seems that permutation equivariance appears to be an advantageous property of the respective neural network model. Our results highlight that a potentially large amount of bailout capital can be saved by utilising models that allow for scenario-dependent allocations and can effectively process the liability matrix -- for example by leveraging permutation equivariance.

For further work, it would be interesting to investigate whether similar theoretical properties can be established for aggregation function based on contagion models with additional default mechanisms, such as fire sales. Besides theoretical challenges due to the non-convexity, this would require to extend the domain of stochastic financial networks to directed heterogeneous graphs with different node and edge types. Subsequently, different neural network architectures might need to be employed for the computation of such systemic risk measures.

\clearpage
\begin{appendix}
\section{Supplementary Theory}

\begin{thm}[Koml{\'o}s Theorem; see \citet{komlos1967generalization} or Lemma 1.70 in \citet{foell}]\label{komlos_thm}
 
Let the convex hull be defined as
\begin{align}\label{def:conv}
    \operatorname{conv}(A) := \left\{\sum_{i=1}^n \alpha_i x_i \,\middle|\, x_i \in A, \alpha_i \geq 0, \sum_{i=1}^n\alpha_i = 1, n \in \N \right\}.
\end{align}
Then, given a sequence $\left(\xi_n\right)$ in $L^0\left(\Omega, \mathcal{F}_0, \mathbb{P} ; \mathbb{R}^d\right)$ such that $\sup _n\left|\xi_n\right|<\infty$ $\mathbb{P}$-almost surely, there exists a sequence of convex combinations
    \begin{align*}
        \eta_n \in \operatorname{conv}\left\{\xi_n, \xi_{n+1}, \ldots\right\}
    \end{align*}
    which converges $\mathbb{P}$-almost surely to some $\eta \in L^0\left(\Omega, \mathcal{F}_0,\mathbb{P} ; \mathbb{R}^d\right)$.
\end{thm}
\end{appendix}

\end{document}